\documentclass[a4paper,11pt]{article}
\pdfoutput=1 
\usepackage{jheppub, bm, color} 
\usepackage[T1]{fontenc}
\usepackage{amssymb,amsfonts,slashed,amsthm,amsmath,graphicx}
\title{Inflation in random Gaussian landscapes}

\author[1]{Ali Masoumi,}
\author[1]{Alexander Vilenkin,}
\author[1,2]{Masaki Yamada}

\affiliation[1]{Institute of Cosmology, Department of Physics and Astronomy, 
Tufts University, Medford, MA  02155, USA}

\affiliation[2]{Department of Physics, Tohoku University, 
Sendai, Miyagi 980-8578, Japan}

\emailAdd{ali@cosmos.phy.tufts.edu}
\emailAdd{vilenkin@cosmos.phy.tufts.edu}
\emailAdd{Masaki.Yamada@tufts.edu}

\def\({\left(}
\def\){\right)}
\def\[{\left[}
\def\]{\right]}

\def\nn{\nonumber \\}

\def\pot{U}
\def\field{\phi}
\def\hs{\zeta}
\def\grad{\eta}
\def\thi{\rho}
\def\Tr{\rm Tr}

\def\lmk{\left(}
\def\rmk{\right)}
\def\lkk{\left[}
\def\rkk{\right]}
\def\dd{{\rm d}}

\newcommand{\beq}{\begin{eqnarray}} 
\newcommand{\eeq}{\end{eqnarray}}

\newcommand{\bel}[1] {\begin{equation}\label{#1}}
\newcommand{\beal}[1] {\begin{eqnarray}\label{#1}}
\newcommand{\be}{\begin{equation}}
\newcommand{\ee}{\end{equation}}
\newcommand{\bea}{\begin{eqnarray}} 
\newcommand{\eea}{\end{eqnarray}}

\newcommand{\expec}[1]{\langle #1 \rangle}

\abstract{We develop analytic and numerical techniques for studying the statistics of slow-roll inflation in random Gaussian landscapes.  As an illustration of these techniques, we analyze small-field inflation in a  one-dimensional landscape. We calculate the probability distributions for the maximal number of e-folds and for the spectral index of density fluctuations $n_s$ and its running $\alpha_s$.  These distributions have a universal form, insensitive to the correlation function of the Gaussian ensemble.  We outline possible extensions of our methods to a large number of fields and to models of large-field inflation.  These methods do not suffer from potential inconsistencies inherent in the Brownian motion technique, which has been used in most of the  earlier treatments.}

\begin{document}
\maketitle
\flushbottom

\section{Introduction} 
\label{sect:Intro}

String theory appears to predict a vast landscape of vacuum states with diverse properties \cite{BoussoPolchinski, Susskind:2003kw}. The string landscape, however, is very complicated and its present understanding is rather limited. The expected number of vacua is exponentially large, so it is not possible to study each of them in detail, and one has to resort to a statistical description.  Early attempts in this direction have been made in Refs. \cite{BoussoPolchinski, Douglas, Denef}.   The approach taken in most of the recent work is to investigate the vacuum statistics in multi-field random potentials with the hope that this will provide insights into the qualitative features of the string landscape. 

One of the key problems that needs to be addressed is to determine the number of vacuum states in the landscape and the distribution of their energy densities.  It has been studied for different kinds of random landscapes, both analytically and numerically, in Refs.~\cite{Easther,Frazer,Bachlechner,Wang,MV,EastherGuthMasoumi}.
The statistics of vacuum decay rates in random landscapes has been investigated in \cite{Masoumi,Paban,MV}.  Another crucial set of problems is to determine the likelihood of slow-roll inflation in the landscape, the expected number of e-folds, and the expected spectrum of density perturbations.  These issues have been explored in Refs.~\cite{Tegmark,Easther,Frazer,Battefeld,Susskind,Yang:2012jf,Marsh:2013qca,Pedro:2013nda,Freivogel:2016kxc,Pedro:2016sli}.  

Numerical studies of large random landscapes are computationally very expensive and become prohibitive when the number of fields is $N\gtrsim 10$.  An interesting method to circumvent this problem was proposed by Marsh {\it et al} \cite{Marsh:2013qca}.  They noted that in order to deduce the inflationary properties of the landscape one only needs to know the potential $U(\phi_1, ..., \phi_N)$ in the vicinity of the inflationary paths.  Furthermore, they conjecture that the evolution of the Hessian matrix $\hs_{ij} = \partial^2 U /\partial\phi_i \partial\phi_j$ along a given path in the landscape is described by a stochastic process that they specify (Dyson Brownian motion).  With this assumption they show that inflation is far less likely than one might expect.  Even if the slow-roll conditions are satisfied in a small patch of the landscape, the slope of the potential tends to rapidly steepen beyond that patch.  This cuts inflation short and induces strong deviations from scale-invariance in the perturbation spectrum.  Similar conclusions have been reached by Freivogel {\it et al} \cite{Freivogel:2016kxc}, who also used the Brownian motion method.  This method, however, has some problematic features \cite{EastherGuthMasoumi}.  Apart from being somewhat {\it ad hoc}, it ignores some important consistency conditions
(e.g., $\partial^2 \hs_{ij} /\partial\phi_k \partial\phi_l = \partial^2 \hs_{kl} /\partial\phi_i \partial\phi_j$ and the constraints from the Morse theory) and correlations (such as the correlation of $\hs_{ij}$ with $U$).\footnote{If there is no correlation between the height of the potential and its second derivative, 
the potential may not be bounded below, which is actually the case in the Brownian motion model.} Moreover, the potential evolved along a closed path does not come back to the original value.  Some other problems with the Brownian motion method have been pointed out in Refs.~\cite{Marsh:2013qca,Wang:2016kzp,Freivogel:2016kxc}.  The status of this method is therefore rather uncertain, and the conclusions it yields for inflation in the landscape should be taken with caution.

A potential problem with large landscape theories has been pointed out by Dine \cite{Dine:2015szg}.
Quantum corrections to the potential get larger as the number of fields $N$ is increased, so the theory may become perturbatively inconsistent when $N$ gets sufficiently large.  We address this issue in Appendix \ref{sec:perturbativity}  for the case of random Gaussian landscapes.  We argue that the problem may not arise in generic models for $N\lesssim 200$ and that much larger values of $N$ may be allowed in axion-type models. 

The main goal of the present paper is to develop  analytic and numerical techniques for calculating probability distributions in random landscapes and outline how they can be used to study the inflationary statistics.
We focus on the simple case of Gaussian landscapes, where the potential is a random Gaussian variable.  
In the next section we introduce random Gaussian potentials and calculate some correlators of the potentials and their derivatives.  In a wide class of inflationary models (the so-called small-field inflation), 
we only need to know the potential in a small vicinity of a critical point (a maximum or an inflection point), and most of the relevant properties are determined if we know the third order Taylor expansion about that point \cite{LindeWestphal}.  We therefore determine, in Section \ref{sec:probability}, the probability distribution for Taylor expansion coefficients up to the third order.  As an illustration, in Section \ref{sec:1D} we apply this distribution to analyze inflation in a one-dimensional landscape.  Possible extensions of the method to higher-dimensional landscapes and to large-field models are outlined in Section \ref{sec:Extensions}.  Our conclusions are summarized in Section \ref{sec:Conclusion}.

\section{Random Gaussian landscapes}\label{sec:Random}

A random Gaussian landscape is defined by a potential $U({\boldsymbol \field})$ in the $N$-dimensional field space $\boldsymbol \field = \{\field_1, \field_2, \ldots, \field_N\}$. The value of the potential at a given point is a random Gaussian variable, but the potential values at different points are correlated. The simplest class of these models have a translationally and rotationally invariant correlator given by
\bel{Correlation}
	\langle \pot (\boldsymbol \field_1) \pot(\boldsymbol \field_2)\rangle=  F (|\boldsymbol \field_1 - \boldsymbol \field_2|)=\frac1{(2\pi)^N}\int d^N k\,P(k) e^{i{\bf k}\cdot (\boldsymbol\field_1-\boldsymbol\field_2)}~, 
\ee
where $k \equiv |\boldsymbol k|$.

We define different moments of the spectral function $P(k)$ as  
\bel{sigmaDef}
	\sigma_{n}^2= \frac1{(2\pi)^N}\int d^N k (k^2)^n P(k)~. 
\ee
A simple choice of the correlation function is 
\be
F(\phi)=U_0^2 e^{-\phi^2/2\Lambda^2},
\label{correlation function}
\ee
with $\Lambda$ playing the role of the correlation length in the landscape.  Then
\be
P(k)= U_0^2 (2\pi\Lambda^2)^{N/2} e^{-\Lambda^2 k^2/2}
\ee
and $\sigma_0^2 =U_0^2$.  In the large-$N$ limit, the higher moments are given by
\be
\sigma_n^2 \approx U_0^2 \left({2N}\over{\Lambda^2}\right)^n.
\ee

Analytic calculations below can be applied to a generic form of the correlation function, 
while we assume Eq.~(\ref{correlation function}) to give some numerical examples.

As we mentioned in the Introduction, for small-field inflation we only need the potential in a small vicinity of a point.  We shall therefore be interested in Taylor expansion coefficients, given by the derivatives of $U(\bf\phi)$, truncating the expansion at the cubic order.  Even after choosing the value of the potential and its derivatives at one point, the values of the potential at other points remain Gaussian variables with a certain mean and standard deviation. In Appendix \ref{sec:singleValued} we calculate these mean values and standard deviations for Taylor expansions of different orders.  We show that the width of the distributions rapidly converges with the order $n$ of the expansion.  This demonstrates the consistency of truncated Taylor expansions for random Gaussian ensembles.

Using \eqref{Correlation} we can derive the correlators between different  derivatives of $U(\bf\phi)$ at a given point. It is easy to show (using spherical symmetry) that even and odd-order derivatives are correlated among themselves, but these two sets are uncorrelated.  For brevity we define 
\beal{derDef}
	 \grad_i = \frac{\partial \pot}{ \partial \field_i},~ \qquad~ \hs_{ij}= \frac{\partial^2 \pot}{\partial \field_i \partial \field_j}, \qquad \thi_{ijk}= \frac{\partial^3 \pot}{ \partial \field_i \partial \field_j \partial \field_k}~.
\eea
The only nonzero correlations are given by
\beal{correlation3}
	\langle \pot, \pot \rangle&=&  \sigma_0^2~, \nn
	\langle \hs_{ij}, \hs_{kl} \rangle&=&  \frac{\sigma_2^2}{N(N+2)}\( \delta_{ij} \delta_{kl} + \delta_{il} \delta_{kj} + \delta_{ik} \delta_{jl}\)~, \nn
	\langle  \pot, \hs_{ij} \rangle&=& -\frac1N \delta_{ij}\sigma_1^2~.\nn
	\langle \grad_i, \grad_j \rangle&=&\frac1N \delta_{ij} \sigma_1^2~, \nn
	\langle \grad_i, \thi_{jkl} \rangle&=&  -\frac{\sigma_2^2}{N(N+2)}\( \delta_{ij} \delta_{kl} + \delta_{il} \delta_{kj} + \delta_{ik} \delta_{jl}\)~,\nn
	\langle \thi_{ijk}, \thi_{lmn} \rangle &=& \frac{\sigma_3^2}{N(N+2)(N+4)} (\delta_{ij} \delta_{kl}\delta_{mn}+ \text{all permutations} )~.
\eea

To further streamline the notation, we define
\beal{varDef}
	E&=\sigma_0^2~, \quad &	A= \frac{\sigma_2^2}{N(N+2)}~, \nn
	B&= -\frac1N \sigma_1^2~,	\qquad &Y=\frac{\sigma_3^2}{N(N+2)(N+4)}~.
\eea
The correlation functions are then given by 
\beal{correlation2}
	\langle \pot, \pot \rangle&=&E~, \nn
	\langle \hs_{ij}, \hs_{kl} \rangle&=&  A\( \delta_{ij} \delta_{kl} + \delta_{il} \delta_{kj} + \delta_{ik} \delta_{jl}\)~, \nn
	\langle  \pot, \hs_{ij} \rangle&=&B \delta_{ij}~, \nn
	\langle \grad_i, \grad_j \rangle&=&-B \delta_{ij}~, \nn
	\langle \grad_i, \thi_{jkl} \rangle&=&  -A\( \delta_{ij} \delta_{kl} + \delta_{il} \delta_{kj} + \delta_{ik} \delta_{jl}\)~,\nn
	\langle \thi_{ijk}, \thi_{lmn} \rangle &=&Y (\delta_{ij} \delta_{kl}\delta_{mn}+ \text{all permutations})~.
\eea

\section{Probability distributions} \label{sec:probability}

For a given set of independent Gaussian variables $\{\alpha_1, \alpha_2, \ldots \alpha_N\}$ with correlations of the form 
\bel{mDef1}
	M_{ij}= \langle \alpha_i \alpha_j \rangle,
\ee
and zero mean, the probability distribution is given by

\bel{Palpha}
	P(\alpha_1, \ldots, \alpha_N)= 
	\sqrt{\frac{ {\rm det} K_{ij} }{ (2\pi)^{N} }}e^{-\frac12 \alpha_i K_{ij} \alpha_j} 
	= \sqrt{\frac{ {\rm det} K_{ij} }{ (2\pi)^{N} }}e^{-Q}~,
\ee
where 
$K=M^{-1}$ and we defined $Q=\frac12 \alpha_i K_{ij} \alpha_j$. Therefore the program of finding the joint distributions of a Gaussian set of variables boils down to finding the inverse matrix $K$ to the correlator matrix $M$. 

Since the Hessian $\hs_{ij}$ and the third derivative $\thi_{ijk}$ are symmetric in their indices, these variables are not independent.  To avoid double (or triple, etc.) counting, we introduce the variables $\hs_{IJ}$ and $\thi_{IJK}$, where the capital Latin letters indicate that the indices are ordered sequentially, so that $I\leq J\leq K$.  We shall use small Latin letters to denote the unrestricted indices, so for example in $\hs_{ij}$ both indices go over the range $[1,N]$.  

According to \eqref{correlation1} the sets of variables $\{ \pot, \hs\}$ and $\{ \grad, \thi\}$ are uncorrelated. Therefore, we only need to calculate the distribution for each set separately. Furthermore, since the Gaussian variables $\alpha_i$ in \eqref{Palpha} are assumed to be independent, we shall use the restricted variables $\hs_{IJ}$ and $\thi_{IJK}$.  This makes the summation over elements tricky and one should be careful about the limits of the summands.  In the end we express our results in terms of the unrestricted variables $\hs_{ij}$, $\thi_{ijk}$, which are much easier to handle.

\subsection{Distribution of $\hs$ and $\pot$}

The correlator matrix $M$ for these variables has the form
\be
	M=\( \begin{array}{cc}
	\langle \pot, \pot \rangle  &  \langle \pot, \hs_{KL} \rangle \\
	 \langle \pot, \hs_{KL} \rangle & \langle \hs_{IJ}, \hs_{KL} \rangle
	\end{array}
	 \)=
	 \( \begin{array}{cc}
	E  &  B \delta_{KL} \\
	 B \delta_{KL} & \langle \hs_{IJ}, \hs_{KL} \rangle
	\end{array}
	 \) .
\ee
The inverse of this matrix has the following elements: 
\beal{firstInverse}
	K_{00}&=&\frac{ (N+2) A}{(N+2)A E -NB^2 }~, \nn
	K_{0,IJ}&=& \frac{B \delta_{IJ}}{N B^2 - (N+2)A E}~, \nn
	K_{IJ,KL}&=& Z \delta_{IJ} \delta_{KL} + \frac1A \delta_{IK}\delta_{JL} - \frac1{2A} \delta_{IL} \delta_{JK}~,
\eea
where 
\beal{secondInverse}
	Z&=&\frac{1}{2A}\frac{B^2- A E}{ (N+2) A E -NB^2 }~.
\eea
The joint probability distribution of the potential and (unrestricted) Hessian, which was derived in \cite{Fyodorov, BrayDean} (using a different notation), is given by
\bel{jointUHess}
	P(\pot, \hs)\propto e^{-Q_{\pot, \hs}}~,
\ee
where
\bel{Q2}
	Q_{\pot, \hs} = \frac12 K_{00} \pot^2 +  \frac{B \pot \Tr \hs}{N B^2 - (N+2)A E} + \frac12Z (\Tr \hs)^2+ \frac1{4A} \Tr \hs^2~.
\ee
Here, $\Tr \hs=\hs_{ii}$, etc.

\subsection{Distribution of $\thi$ and $\grad$}

To our knowledge, the distribution of $\thi$ and $\grad$ has not yet been discussed in the literature.
The correlation structure of these variables gives the following matrix $M$:
\bel{thirdOrderM}
	M=\( \begin{array}{cc}
	\langle \grad_I, \grad_J \rangle  &  \langle \grad_I,\thi_{JKL} \rangle \\
	 \langle \grad_I, \thi_{JKL} \rangle & \langle \thi_{IJK}, \thi_{LMN} \rangle
	\end{array}
	 \)=
	 \( \begin{array}{cc}
	-B \delta_{IJ}  &   -A (\delta_{IJ} \delta_{KL}+ \text{perm}) \\
	 -A (\delta_{IJ} \delta_{KL}+ \text{perm})&Y (\delta_{IJ} \delta_{KL}\delta_{MN}+\text{perm}) 
	\end{array}
	 \).
\ee

The derivation of the inverse matrix $K$ in this case requires somewhat lengthy algebra; we give it in Appendix \ref{sec:InverseMatrix}.  The resulting joint distribution of first and third derivatives is given by
\be
	P(\eta_l, \rho_{ijk})= e^{-Q}~,
\ee
where 
\be
	Q=\frac12 L \eta_i \eta_i+  X \eta_i \rho_{ijj}+ \frac12 C_1 \rho_{iik}\rho_{jjk} + \frac1{12} C_8 \rho_{ijk}\rho_{ijk}~
\ee
with
\bea
	C_1 &=& -\frac{A^2+B Y}{2 Y \left(A^2 (N+2)+B (N+4) Y\right)}\nn
	C_8&=& \frac1Y~, \nn
	L&=&-\frac{(N+4) Y}{A^2 (N+2)+B (N+4) Y} ~,\nn
	X&=& -\frac{A}{A^2 (N+2)+B (N+4) Y}~.
\eea

\section{Inflation in a $1D$ landscape} \label{sec:1D}

To illustrate the application of the above distributions, we shall now apply them to analyze inflation in a one-dimensional landscape.  We define the slow-roll parameters as 
\beq
 \epsilon_s = \frac{1}{2} \lmk \frac{U'}{U} \rmk^2,
 \label{epsilon}
 \eeq
 \beq
 \eta_s = \frac{U''}{U}, 
\label{eta}
\eeq
The necessary conditions for slow-roll inflation are $\epsilon_s, \eta_s\ll 1$.  

We shall focus on small-field inflation, when the correlation length of the potential is $\Lambda \ll 1$.  
The typical values of the slow-roll parameters at a randomly chosen point in the landscape are then $\epsilon_s \sim \eta_s \sim \Lambda^{-2} \gg 1$.  Inflation can thus occur only in rare regions where $U'$ and $U''$ are unusually small.  On the other hand, $U'''$ needs not be particularly small and will typically be of the order $U''' \sim U \Lambda^{-3}$.  The typical range of the inflaton field $\phi$ where the slow-roll conditions hold can then be estimated from $|U'''|\Delta\phi \sim U$, or
\beq
\Delta\phi \sim \Lambda^3 \ll 1.
\label{range}
\eeq
This range is much smaller than the Planck scale, which is the defining property of small-field inflation.

The typical change of the potential in the field range $\Delta\phi$ is 
\beq
\Delta U \approx U' \Delta\phi +\frac{1}{2} U'' (\Delta\phi)^2 +\frac{1}{3!} U''' (\Delta\phi)^3 + \dots \ll U\Lambda^3.
\label{change}
\eeq
Hence, $\Delta U/U \ll \Lambda^3 \ll 1$.  This is another characteristic property of small-field inflation: the potential remains nearly constant in the entire slow-roll region.

In small-field models, inflation can occur near a maximum of the potential (hilltop inflation) or near an inflection point.  As mentioned above, the main features of small-field inflation can usually be captured if we know the potential to cubic order in Taylor expansion. In the $1D$ case the probability distributions for the expansion coefficients are greatly simplified.
The only nonzero correlations in this case are 
\beal{correlation1}
	&\langle \pot, \pot \rangle=  \sigma_0^2~, &\quad		\langle \hs,\hs \rangle= \sigma_2^2~,	 \quad\langle  \pot, \hs\rangle= -\sigma_1^2~,\nn
	&\langle \grad, \grad \rangle= \sigma_1^2~,&\quad 	\langle \grad, \thi \rangle=  -\sigma_2^2~,\quad \langle \thi, \thi \rangle = \sigma_3^2~.
\eea
The joint distribution of $\{\pot, \grad, \hs, \thi\}$ is given by 
\be
	P(\pot, \grad, \hs, \thi) = P_1(\pot, \hs) P_2(\grad, \thi)~,
\ee
where 
\bea
	P_1(\pot, \hs) &=& A_1 \exp \[-\frac12\( \frac{ \sigma _2^2 \pot^2+2 \sigma _1^2 \hs  \pot+\sigma _0^2\hs ^2}{\sigma _0^2 \sigma _2^2- \sigma _1^4 }\)\]~,\nn \nn
	P_2(\grad, \thi)&=& A_2 \exp{\[-\frac12\( \frac{\sigma _3^2\grad^2 +2 \sigma _2^2 \grad\thi+\sigma _1^2 \thi^2}{\sigma _1^2 \sigma _3^2-\sigma _2^4}\)\]}~.
	\label{P_2}
\eea
The normalization constants $A_1$ and $A_2$ are given by 
\bea
        A_1 &=& (2\pi)^{-1} \left(\sigma_0^2 \sigma_2^2 -\sigma_1^4 \right)^{-1/2} ~, \nn \nn
        A_2 &=& (2\pi)^{-1} \left(\sigma_1^2 \sigma_3^2 -\sigma_2^4 \right)^{-1/2} .
        \label{A_2}
\eea

 We shall now study the statistical properties of $1D$ landscapes using the methods developed in Refs. \cite{Rice,Bardeen}.

\subsection{Numbers of extrema and of inflection points}

For any specific realization of a landscape $U(\phi)$ in the range $|\phi|\leq L$, the number of points where the $k$'th derivative of the potential vanishes can be expressed as
\be
N_k=\int_{-L}^L d\field \delta(\pot^{(k)}) |\pot^{(k+1)}|~,
\label{N_k}
\ee
Because of the $\delta$-function the contributions come only from the points where
the $k$'th derivative vanishes,  and the $(k+1)$'th derivative ensures that each of these contributions is equal to unity.  To get the ensemble average of $N_k$, we need the joint distribution of the $k$'th
and $(k+1)$'th derivatives. Hence, we get
\be
\expec{N_k}=\int_{-\infty}^\infty\int_{-\infty}^\infty \int_{-L}^L d\pot^{(k)}d\pot^{(k+1)} d\field P\(\pot^{(k)},\pot^{(k+1)}\)   \delta(\pot^{(k)}) |\pot^{(k+1)}|~.
\label{avl N_k}
\ee
For a Gaussian landscape, $\pot^{(k)}$ and $\pot^{(k+1)}$ are uncorrelated, so the distribution factorizes, 
\be
P\(\pot^{(k)}, \pot^{(k+1)}\) = \frac1{2\pi \sigma_k \sigma_{k+1}}\exp\(-\frac{1}{2\sigma_k^2}(\pot^{(k)})^2-\frac{1}{2\sigma_{k+1}^2}(\pot^{(k+1)})^2\).
\ee
This leads to 
\be
\expec{N_k}= (2L) \frac{\sigma_{k+1}}{\pi \sigma_{k}}~.
\label{expec N_k}
\ee
As can be expected from translational invariance of the distribution, this is proportional to $2L$. We can define the density of these points per unit of length in field space by dividing by $2L$, 
\bel{dDerDensity}
n_k = \frac{\sigma_{k+1}}{\pi\sigma_k}~.
\ee

We are interested in the densities of extrema and of inflection points, which are given respectively by
\bel{n1}
n_1 = \frac{\sigma_{2}}{\pi\sigma_1}~,
\ee
\bel{n2}
n_2 = \frac{\sigma_{3}}{\pi\sigma_2}~.
\ee
Since $\sigma_k\sim U_0/\Lambda^k$, we have $n_1\sim n_2 \sim \Lambda^{-1}$.  The density of maxima is $n_1/2$.  The relative number of inflection points and extrema is given by
\be
\frac{n_2}{n_1}=\frac{\sigma_1\sigma_3}{\sigma_2^2}.
\label{n_2/n_1}
\ee
Since there should be at least one inflection point between any adjacent extrema, we expect that $n_2/n_1>1$, which implies $\sigma_1\sigma_3>\sigma_2^2$.  And indeed this condition must be satisfied, since otherwise the distribution (\ref{P_2}) is not normalizable.

In the following subsections we shall calculate the probability distributions for the maximal number of e-folds $N_{tot}$ and for the spectral index of density perturbations $n_s$, both for inflection point and hilltop inflation.   Given a function $f(U, \grad, \hs, \thi)$, its conditional probability distributions at hilltops and inflection points can be evaluated as
\beq
 P_k (f) &=&  
 \frac{1}{\expec{N_k}} \int \dd \phi \dd U \dd \grad \dd \hs \dd \thi P(U, \grad, \hs, \thi) 
 \delta \lmk U^{(k)} \rmk \left\vert U^{(k+1)} \right\vert 
\delta \lkk f - f(U, \grad, \hs, \thi) \rkk .
\label{P(f)}
\eeq
Here, $f$ can be either $N_{\rm tot}$ or $n_s$.  The integral with respect to $\phi$ together with 
$\delta \lmk U^{(k)} \rmk \left\vert U^{(k+1)} \right\vert $ 
finds hilltops ($k=1$) or inflection points ($k=2$).  The factor $\expec{N_k}$, defined by Eq.~(\ref{avl N_k}), comes from the normalization of the probability. 
Note that the integral with respect to $\phi$ gives a factor of $2L$, 
which is an arbitrary parameter and disappears in the final expression 
due to the cancellation with $1/\expec{N_k}$ [see Eq.~(\ref{expec N_k})].

\subsection{Inflection point inflation} 
\label{sec:inflection}

To consider inflection point inflation, first suppose that 
the second derivative $U''$ vanishes at $\phi = 0$ and expand the potential at that point:
\beq
 U(\phi) = U + \grad \phi + \frac{\thi}{3!} \phi^3, 
 \label{cubic}
\eeq
Note that $\grad \thi$ should be positive in this case, since otherwise the potential has a local maximum near $\phi=0$.  
Inflation models with a cubic potential (\ref{cubic}) have been extensively studied in the literature \cite{Baumann,LindeWestphal}.

Inflation can occur near $\phi=0$ only if $|U'|,|U''| \ll U$.  It ends when the field value reaches 
$\phi_{\rm end}\sim U/ \rho$, at which the slow-roll conditions fail. 
The total number of inflationary e-foldings is then bounded by\footnote{The actual number of e-folds depends on the initial conditions for the slow roll, specifically on the initial value of $\phi$.}
\beq
 N_{\rm tot} \approx \int_{-\infty}^\infty d\phi\frac{U(\phi)}{U'(\phi)} 
 = \pi \sqrt{2} \frac{U}{\sqrt{\grad \thi}}, 
 \label{Ntot}
\eeq
where we assume $N_{\rm tot} \gg 1$ and $\thi \gg U$. 
The latter condition guarantees that inflation is of the small-field type, that is, we can use $U(\phi) \simeq U$  in the slow-roll region. Note that the typical value of $\rho$ is 
of order $U / \Lambda^3$, so that $\thi \gg U$ is typically satisfied when $\Lambda \ll 1$. 
The probability distribution for $N_{\rm tot}$ for the inflection point inflation 
can be calculated from [see Eq.~(\ref{P(f)})] 
\beq
 P_{\rm inflection} (N_{\rm tot}) = 
 \frac{1}{\expec{N_2}} \int \dd \phi \dd U \dd \grad \dd \hs \dd \thi P(U, \grad, \hs, \thi) 
 \delta \lmk \hs \rmk \left\vert \thi \right\vert 
\delta \lmk N_{\rm tot} - \pi \sqrt{2} \frac{U}{\sqrt{\grad \thi}} \rmk, 
\nonumber\\
\eeq
where the integral with respect to $\phi$ together with $\delta \lmk \hs \rmk \left\vert \thi \right\vert $ 
finds the inflection points. 

Integrating out the delta-functions and using Eq.(\ref{expec N_k}) for $\expec{N_2}$, we have
\beq
 P_{\rm inflection} (N_{\rm tot}) &=& \frac{4 \pi^3 A_1 A_2}{N_{\rm tot}^3} 
 \frac{\sigma_2}{\sigma_3}  \int \dd U \, 
 U^2 {\rm exp} \lkk - \frac{1}{2} \lmk \frac{\sigma_2^2}{\sigma_0^2 \sigma_2^2 - \sigma_1^4} \rmk U^2 \rkk 
 \nonumber\\
 &&\quad \times \int \dd \thi \, {\rm exp} 
 \lkk - \frac{1}{2 \lmk \sigma_1^2 \sigma_3^2 - \sigma_2^4 \rmk} 
 \lmk \sigma_1^2 \thi^2 + \frac{4 \pi^2 \sigma_2^2 U^2}{N_{\rm tot}^2} 
 + \frac{\sigma_3^2}{\thi^2} \lmk \frac{2 \pi^2 U^2}{N_{\rm tot}^2} \rmk^2  \rmk \rkk, 
 \nonumber\\
\label{PN}
\eeq
where $A_1$ and $A_2$ are given by Eq.~(\ref{A_2}).  
With $\sigma_n^2 \sim U_0^2 / \Lambda^{2n}$, the second term in the parentheses in Eq.~(\ref{PN}) makes a negligible contribution to the exponent $(\sim \Lambda^4 N_{\rm tot}^{-2})$, while the first and third terms provide effective lower and upper cutoffs for the integral with respect to $\thi$. 
Note also that our assumption that $\thi \gg U$ gives a stronger lower cutoff for the integral.
 In any case, the integral is dominated by the upper cutoff, so we can approximate it by a Gaussian integral with only the first term in the exponent.  This gives
\beq
 P_{\rm inflection} (N_{\rm tot}) \simeq 
 \frac{\pi^2}{N_{\rm tot}^3} \frac{\sigma_0^2 \sigma_2^2 - \sigma_1^4}{\sigma_1 \sigma_2^2 \sigma_3} 
 = \frac{8\pi^2}{9 \sqrt{15}} \frac{\Lambda^4}{N_{\rm tot}^3} , 
\label{PNtot}
\eeq
where in the last step we used the correlation function (\ref{correlation function}). 
Similar calculations of the distribution for $N_{\rm tot}$ have been performed earlier in the literature \cite{Susskind,Agarwal,Yang:2012jf}.  However, these authors did not have a well motivated distribution for $\grad$ and $\thi$ and had to make some {\it ad hoc} choices.  The distributions they obtained differ from ours either in the exponent of $N_{\rm tot}$ or in the prefactor.   Note that the $N_{\rm tot}^{-3}$ dependence in Eq.~(\ref{PNtot}) applies for any form of the correlation function $F$ in (\ref{Correlation}).

\begin{figure}[t] 
   \centering
   \includegraphics[width=4in]{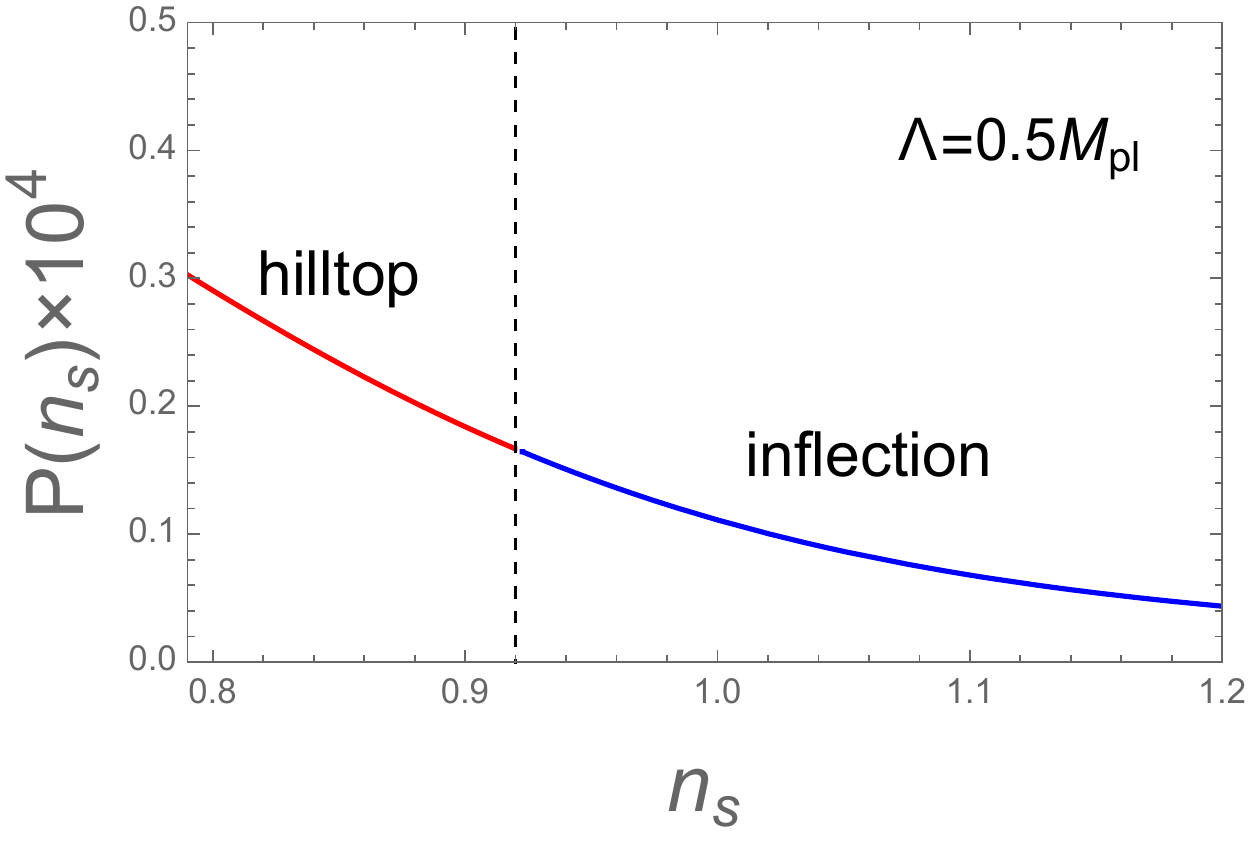} 
   \caption{
   Analytic probability distribution of the spectral index for the case of hilltop inflation (red lines) 
   and inflection point inflation (blue line).   We use the correlation function (\ref{correlation function}) with $\Lambda = 0.5$ in this example.  The distribution is independent of $U_0$ and the dependence on $\Lambda$ is trivial ($\propto \Lambda^4$).}
   \label{fig:analytic}
\end{figure}

The spectral index of density perturbations $n_s$ and its running $\alpha_s$ for inflection point inflation 
can be expressed as \cite{Baumann}
\beq
 &&n_s \simeq 1 - \frac{4\pi}{N_{\rm tot}} {\rm cot} \lmk \frac{\pi N_e}{N_{\rm tot}} \rmk 
 \label{n_s for inflection}
 \\
 &&\alpha_s = \frac{\dd n_s}{\dd \ln k} 
 = - \frac{\dd n_s}{\dd N_e} 
 \simeq - \frac{4 \pi^2}{N_{\rm tot}^2} 
 {\rm sec}^2 \lmk \pi \frac{N_e}{N_{\rm tot}} \rmk, 
 \label{running for inflection}
\eeq
where $N_e$ ($\simeq 50$-$60$) is the e-folding number at which the CMB scale leaves the horizon. 
These expressions depend only on the combination $N_{\rm tot} \propto U/\sqrt{\grad \thi}$, 
and thus they are rigidly correlated with one another.  It follows from Eq.~(\ref{n_s for inflection}) that $n_s$ is greater than $1-4/N_e \approx 0.92$.  This value is approached in the limit $N_{\rm tot}\to\infty$.  In the same limit, $\alpha_s \to -4/N_e^2 \approx -0.0064$ (for $N_e=50$).   The probability distribution for $n_s$ can be found from
\beq
 \tilde{P}_{\rm inflection}(n_s) = P_{\rm inflection}(N_{\rm tot}(n_s)) \left\vert \frac{\dd n_s}{\dd N_{\rm tot}} \right\vert^{-1}. 
\label{Pns}
\eeq
This is shown in Fig.~\ref{fig:analytic} as a blue line, 
where we multiply Eq.~(\ref{Pns}), which is a conditional probability for inflection point inflation, 
by a factor of $n_2 / (n_1 + n_2)$ [see Eq.~(\ref{n_2/n_1})] in order to compare 
this with the result for hilltop inflation discussed in the next subsection. 

 The probability that slow-roll inflation with $N_{\rm tot} \gtrsim N_e$ occurs at a randomly chosen inflection point can be estimated as
\beq
 P \lmk {N_{\rm tot} \gtrsim N_e} \rmk 
 &\approx& \int_{N_e}^\infty \dd N_{\rm tot} P_{\rm inflection} (N_{\rm tot}) 
 \\
 &\simeq& 
\frac{\pi^2}{2N_e^2} \frac{\sigma_0^2 \sigma_2^2 - \sigma_1^4}{\sigma_1 \sigma_2^2 \sigma_3}. 
\eeq

 The observed value of the spectral index, $n_s \simeq 0.97$, corresponds to $N_{\rm tot} \simeq 122 \equiv N_*$.  To see how typical this value is, we consider the set of inflection points with $N_{\rm tot}>N_e$ and
ask what fraction of these points have $N_{\rm tot}> N_*$ (or equivalently $n_s < 0.97$).  This is given by
\beq
 \frac{1}{ P \lmk {N_{\rm tot} \gtrsim N_e} \rmk } 
 \int_{N_*}^\infty \dd N_{\rm tot} P(N_{\rm tot}) 
 \simeq \lmk \frac{N_e}{N_*} \rmk^2 \simeq 0.17,
\eeq
indicating that the observed value is reasonably typical.

\subsection{Hilltop inflation} 
\label{sec:inflection}

Hilltop inflation occurs near a local maximum of $U(\phi)$.  We can then use the expansion
\beq
U (\phi) = U + \frac{\hs}{2!} \phi^2 + \frac{\thi}{3!} \phi^3, 
 \label{U for hilltop}
\eeq
with $\phi=0$ at the maximum. In this case, inflation is eternal near
the maximum \cite{AV83,AVTopological, LindeTopological}, so $N_{\rm tot}=\infty$.\footnote{Eternal inflation is also possible in the vicinity of an inflection point, when
$\grad \lesssim 0.1 U_0^{3/2}$.  Here we disregard this possibility.  This may be justified when $\Lambda, U_0\ll 1$ and regions of such small $\grad$ are very rare.}

The spectral index and its running are given by~\cite{LindeWestphal}
\beq
 &&n_s \simeq 1 - 2 \frac{\hs}{U} {\rm coth} 
 \lmk \frac{\hs N_e}{2 U} \rmk
 \label{n_s in hilltop}
 \\
 &&\frac{\dd n_s}{\dd \ln k} 
 = - \frac{\dd n_s}{\dd N_e} 
 \simeq - \lmk \frac{\hs }{U} \rmk^2 
 {\rm sech}^{2} \lmk  \frac{\hs N_e}{2 U} \rmk. 
\eeq
These results can be obtained from Eqs.~(\ref{n_s for inflection}) and (\ref{running for inflection}) via the analytic continuation to negative $\grad \thi$ by replacing $\grad \thi \to - \hs^2 /2$. 
These expressions depend only on the combination $\hs / U$, 
and thus they are again completely correlated with each other. 
Note that $n_s$ is less than about $0.92$ in this case, 
so that this is complementary to the case of inflection point inflation. 

The probability distribution of spectral index can be directly calculated from 
\beq
 P_{\rm hilltop} (n_s) &=&  
 \frac{1}{\expec{N_1}} \int \dd \phi \dd U \dd \grad \dd \hs \dd \thi P(U, \grad, \hs, \thi) 
 \delta \lmk \grad \rmk \left\vert \hs \right\vert 
\delta \lkk n_s - 1 + 2 \frac{\hs}{U} {\rm coth}
 \lmk \frac{\hs N_e}{2 U} \rmk \rkk. 
 \nonumber\\
\eeq
The delta functions are eliminated by the integrals with respect to $\grad$ and $\hs$. 
Then using $\int \dd \phi / \expec{N_1} = \pi \sigma_1 / \sigma_2$, 
we obtain 
\beq
  P_{\rm hilltop} (n_s) &=&  
 \frac{\pi \sigma_1}{\sigma_2} \int \dd U \dd \thi P(U, \grad = 0, \hs = \hs (n_s), \thi) 
 U^2
 \frac{\left\vert \hs \right\vert}{U} 
 \frac{1}{U \left\vert \frac{\dd n_s}{\dd \hs} \right\vert} 
 \\
 &\simeq& 
 \frac{\sigma_0^2 \sigma_2^2 - \sigma_1^4}{4 \sigma_2^4} 
  \frac{\left\vert \hs \right\vert}{U} 
 \frac{1}{U \left\vert \frac{\dd n_s}{\dd \hs} \right\vert}, 
\label{Pns2}
\eeq
where the combinations $\hs / U$ and $U \left\vert \frac{\dd n_s}{\dd \hs} \right\vert$ 
should be expressed in terms of $n_s$ using Eq.~(\ref{n_s in hilltop}). 
In the second line, we assume $\zeta \ll  U_0 / \Lambda^2$ to neglect the exponential factor, 
which is justified for $| 1- n_s | \ll 1$. 
The result is shown in Fig.~\ref{fig:analytic} 
as a red line, 
where we multiply Eq.~(\ref{Pns2}) 
by a factor of $n_1 / (n_1 + n_2)$ [see Eq.~(\ref{n_2/n_1})] in order to compare 
this with the result for inflection point inflation. 
These results perfectly match at $n_s \simeq 0.92$.

\subsection{Numerical results}

To check the analytic results described above, and also to check the accuracy of the expansion of the potential to the third power (\ref{cubic}), we performed some numerical simulations.
We consider two examples, where the Taylor expansion is truncated at the third or seventh order, 
that is, the inflaton potential is assumed to be 
\beq
 U(\phi) = 
 U^{(0)} + U^{(1)} \phi + \frac{U^{(2)} }{2!} \phi^2 
 + \dots + \frac{U^{(n_{\rm max})}}{n_{\rm max}!} \phi^{n_{\rm max}}, 
\eeq
with $n_{\rm max} = 3$ or $7$.  The correlators of the potential and its derivatives are given by 
\bel{oneFieldCorrelation}
	\expec{\pot^{(m)}(\field) \pot^{(n)}(\field)} =i^n (-i)^m \sigma_{(n+m)/2}^2 
\ee
and are nonzero only when the sum of $n$ and $m$ is an even integer.  For $n_{\rm max}=3$, the probability distribution for $U^{(i)}$ is given by Eq.~(\ref{P_2}), where $U \equiv U^{(0)}$, $\grad \equiv U^{(1)}$, $\hs \equiv U^{(2)}$, and $\thi \equiv U^{(3)}$.  

For $n_{\rm max}=7$, the sets $\alpha=\{\pot^{(0)},\pot^{(2)},\pot^{(4)},\pot^{(6)} \}$ and $\beta=\{\pot^{(1)},\pot^{(3)},\pot^{(5)},\pot^{(7)} \}$ are uncorrelated. Let us define the matrices $M^{\alpha}$ and $M^{\beta}$ as 
\be
	M^{\alpha}_{ij}= \expec {\alpha_i, \alpha_j}~, \qquad M^{\beta}_{ij}= \expec {\beta_i, \beta_j}~.
\ee
Using \eqref{oneFieldCorrelation} we get
\be
	M^{\alpha}=\left(
\begin{array}{cccc}
 \sigma ^2_0 & -\sigma ^2_1 & \sigma ^2_2 & -\sigma ^2_3 \\
 -\sigma ^2_1 & \sigma ^2_2 & -\sigma ^2_3 & \sigma ^2_4 \\
 \sigma ^2_2 & -\sigma ^2_3 & \sigma ^2_4 & -\sigma ^2_5 \\
 -\sigma ^2_3 & \sigma ^2_4 & -\sigma ^2_5 & \sigma ^2_6 \\
\end{array}
\right), \qquad 
M^{\beta}=\left(
\begin{array}{cccc}
 \sigma ^2_1 & -\sigma ^2_2 & \sigma ^2_3 & -\sigma ^2_4 \\
 -\sigma ^2_2 & \sigma ^2_3 & -\sigma ^2_4 & \sigma ^2_5 \\
 \sigma ^2_3 & -\sigma ^2_4 & \sigma ^2_5 & -\sigma ^2_6 \\
 -\sigma ^2_4 & \sigma ^2_5 & -\sigma ^2_6 & \sigma ^2_7 \\
\end{array}
\right)~.
\ee
The inverses of these matrices are easy to calculate using Mathematica, although not very compact to write here.  We can then use  Eq.~(\ref{Palpha}) to find the probability distributions for the variables $\alpha_j$ and $\beta_j$.  Using these distributions, we generated $10^{10}$ realizations of the potential and selected those that satisfy the slow-roll conditions and have $N>N_e$ e-foldings in the slow-roll regime.  We then calculated the distributions for $n_s$ and $\alpha_s$.

The spectral index and its running can be expressed as
\beq
 &&n_s = 1- 6 \epsilon_s + 2 \eta_s, 
 \\
 &&\frac{{\rm d} n_s}{{\rm d} {\rm ln} k } 
 = - \frac{\dd n_s}{\dd N_e} 
 = - 24 \epsilon_s^2 + 16 \epsilon_s \eta_s - 2 \xi_s.,
\eeq
where $\epsilon_s$ and $\eta_s$ are given by Eqs.~(\ref{epsilon}),(\ref{eta}) and we have defined 
\beq
\xi_s = \frac{U' U'''}{U^2}. 
\eeq

In each realization of the potential that has a slow-roll region with $\epsilon_s,\eta_s<1$, 
we first find a point ${\bar \phi}$ in the field space at which the slow-roll conditions fail ($\epsilon_s =1$ or $\eta_s =1$) and then look for a point $\phi_e$ at which the number of e-foldings is $N_e \simeq 50$, 
\bel{efoldNum}
 N_e \simeq \int_{\bar \phi}^{\phi_e} \frac{U}{U'} {\rm d} \phi = 50. 
\ee
Note that the results do not depend on the overall height of the potential $U_0$.  Among the $10^{10}$ realizations, only $\mathcal{O}(10^{5})$ have sufficient number of e-foldings.

\begin{figure}[t] 
   \centering
   \includegraphics[width=2.5in]{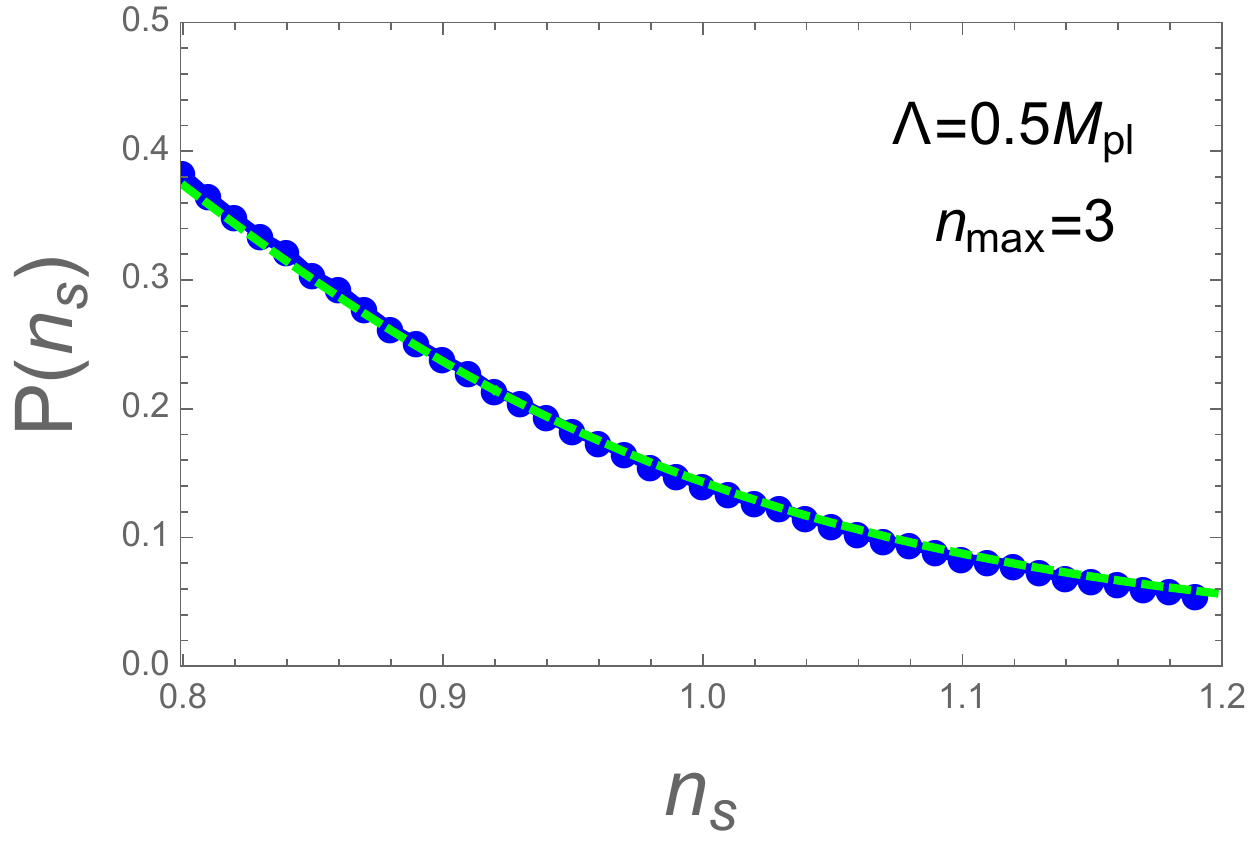} 
   \quad
   \includegraphics[width=2.5in]{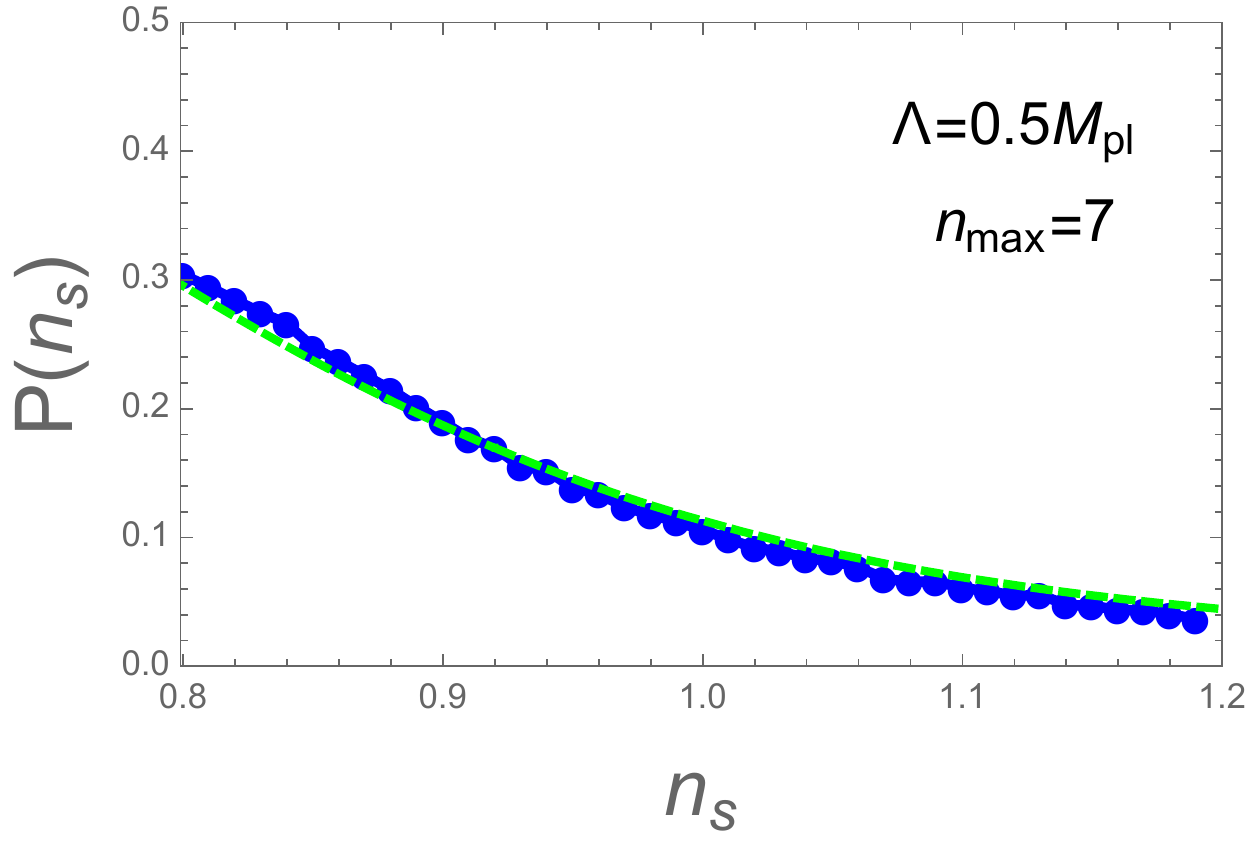} 
   \caption{
   Distribution of values of the spectral tilt $n_{s}$ for numerically generated ensembles with  $n_{\rm max} = 3$ (left panel) 
   and $n_{\rm max} = 7$ (right panel). We only kept the samples which had a sufficient number of e-folds.
   The green dashed line is the analytic distribution given by Eqs.~(\ref{Pns}) and (\ref{Pns2}). 
   The normalization is arbitrary. 
   }
   \label{fig:histogram}
\end{figure}

Figure~\ref{fig:histogram} is a histogram of the number of realizations 
as a function of $n_s$ for the case of $\Lambda = 0.5 M_{\rm pl}$ with 
$n_{\rm max}=3$ (left panel) and $n_{\rm max}=7$ (right panel). The relatively large value of $\Lambda$ has been chosen because otherwise realizations having the required number of e-foldings are exceedingly rare (see Eq.~(\ref{PNtot})).  The plots are in a good agreement with the analytic results.

\begin{figure}[t] 
   \centering
   \includegraphics[width=2.5in]{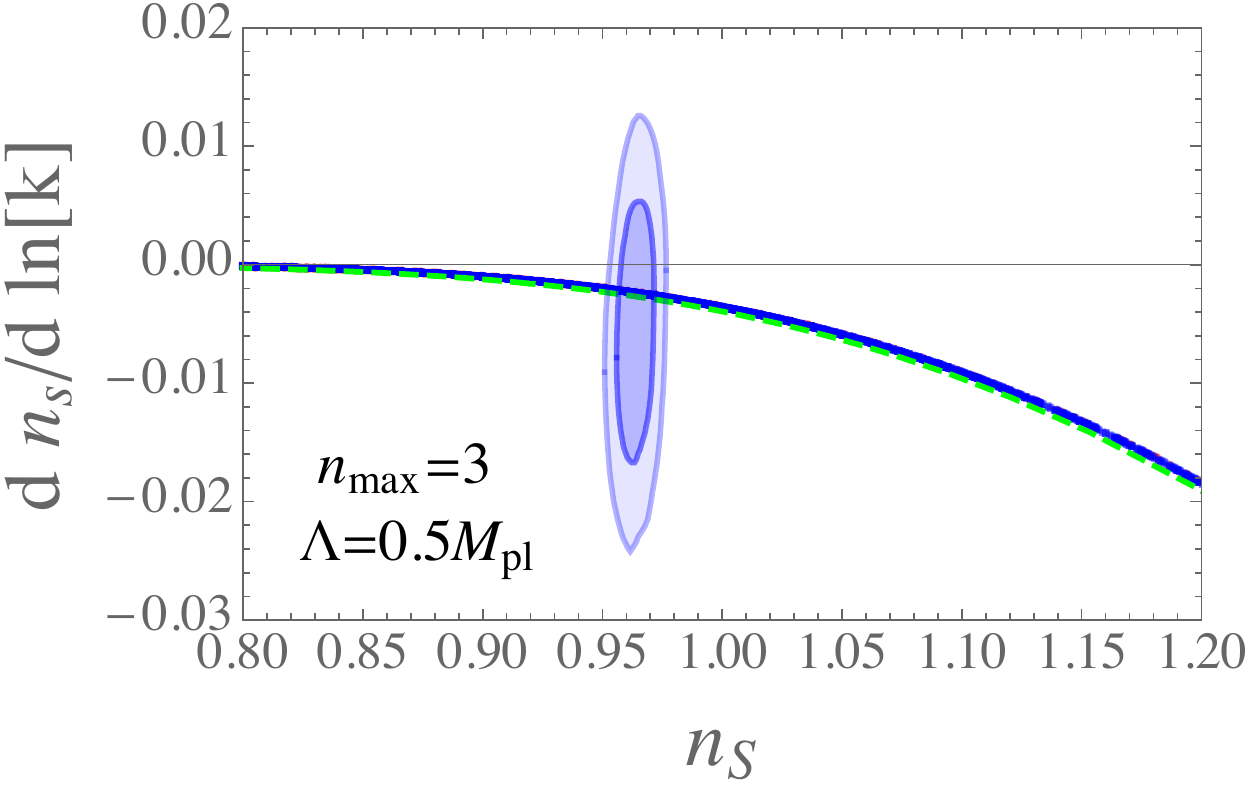} 
   \quad
   \includegraphics[width=2.5in]{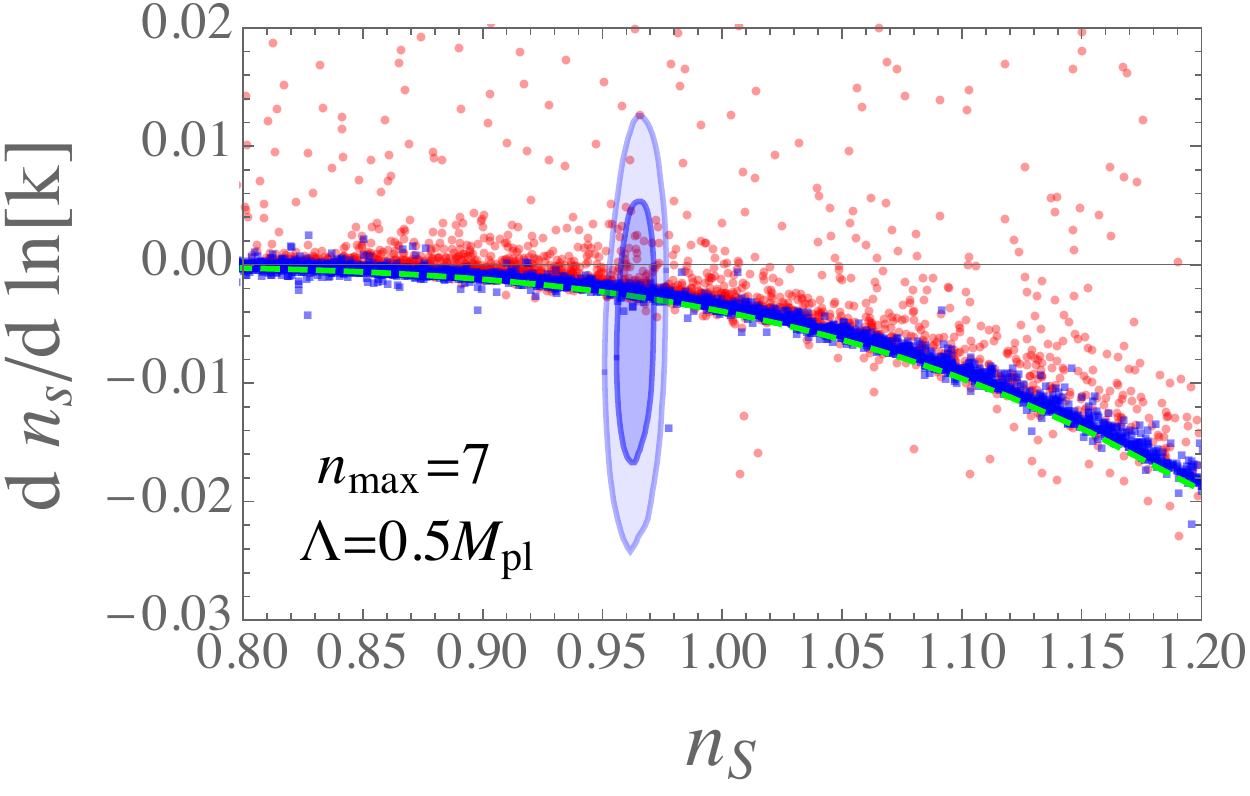} 
   \caption{
   Relation between the running of the spectral index ${\rm d} n_s / {\rm d} {\rm ln} k$ 
   and the spectral index $n_s$ 
   for the case of $\Lambda = 0.5 M_{\rm pl}$ with $n_{\rm max} = 3$ (left panel) 
   and $n_{\rm max} = 7$ (right panel). 
Each blue (red) dot represents one realization that satisfies $|U'''|/U> 10$ ($|U'''|/U< 10$). 
   The green dashed line is the prediction of the analytic formula. 
   The light (darker) blue-shaded region is $68\%$ (95\%) CL of the Planck result~\cite{Ade:2015lrj}. 
   }
   \label{fig:running}
\end{figure}

We also show the relation between the spectral index $n_s$ 
and its running $\alpha_s = {\rm d} n_s / {\rm d} {\rm ln} k$ in Fig.~\ref{fig:running}. 
The agreement with the analytic result is very close for the case of $n_{\rm max} = 3$. 
For $n_{\rm max}=7$, the higher order terms in the inflaton potential introduce some scatter in the data, as we can see in the right panel of Fig.~\ref{fig:running}.  
These aberrations occur in realizations violating the condition for small-field inflation, $|U'''|/U \gg 1$.  To illustrate this point, we used red (blue) dots to represent realizations with $|U'''|/U< 10$ ($|U'''|/U> 10$).  Despite the relatively large number of red dots (about $ 34\%$), most of them lie close to the analytic curve: the fraction of data points with values of $\alpha_s$ deviating from the analytic result by more than 0.002 is only about $6\%$.  We expect the number of realizations violating the small-field condition $|U'''|/U \gg 1$  to decrease for smaller values of $\Lambda$.  
We conclude that expansion up to a cubic term is well justified and is likely to become more accurate for small $\Lambda$.

 Also indicated in Fig.~\ref{fig:running} are the Planck satellite results for $\alpha_s$ and $n_s$~\cite{Ade:2015lrj}, with the lighter and darker blue-shaded regions indicating the 68\%  and 95\% CL values, respectively.  We thus see that the observed values are consistent with a random Gaussian multiverse.

\section{Extensions to $N>1$ and to large-field inflation}\label{sec:Extensions}

 The methods we developed here can be straightforwardly applied to models of small-field inflation in higher-dimensional landscapes ($N>1$).  We expect that Taylor expansion of the potential up to third order should still give accurate results.  Then one can use the distributions we found in Section \ref{sec:probability} to study the statistics of the inflationary parameters. 

In a large-field landscape with $\Lambda > 1$, the slow-roll range is typically $\Delta\phi >\Lambda$
(see Eq.~(\ref{range})), and 
such a truncated Taylor expansion would not generally be applicable.  Expanding to higher orders or using Fourier expansion is computationally costly for a large number of fields and becomes prohibitive for $N\gtrsim 10$.  A possible way to get around this difficulty is to Taylor-expand the potential in little patches along the inflationary track, as in the Dyson Brownian motion method \cite{Marsh:2013qca}.  However, as we mentioned in the Introduction, this method involves some {\it ad hoc} assumptions and suffers from some potential inconsistencies.  We shall now outline how the idea of following the inflationary track can be adopted within our approach.  Our procedure respects all the correlations and does not make any {\it ad hoc} assumptions about the evolution of the Hessian and other quantities.

\subsection{Following the inflationary track} \label{subsec:alongAPath}

We start at some point $\field_0$ and expand the potential to $n$-th order,  
\be
     \pot({\bf\field}_0 +\delta {\bf\field})= \pot_0 + \sum _{m=1}^n \sum_{j_i=1}^N A^{(0)}_{j_1, \ldots j_m} \delta \field_{j_1} \delta \field_{j_2}\ldots \delta \field_{j_m}~.
\label{Taylor}
\ee
The values of $U_0$ and $A^{(0)}_{j_1, \ldots j_m}$ are taken from the Gaussian ensemble, and we are interested only in the subset of this ensemble where the slow-roll conditions are satisfied at ${\bf\phi}={\bf\phi}_0$.
\begin{figure}[t] 
   \centering
   \includegraphics[width=3.5in]{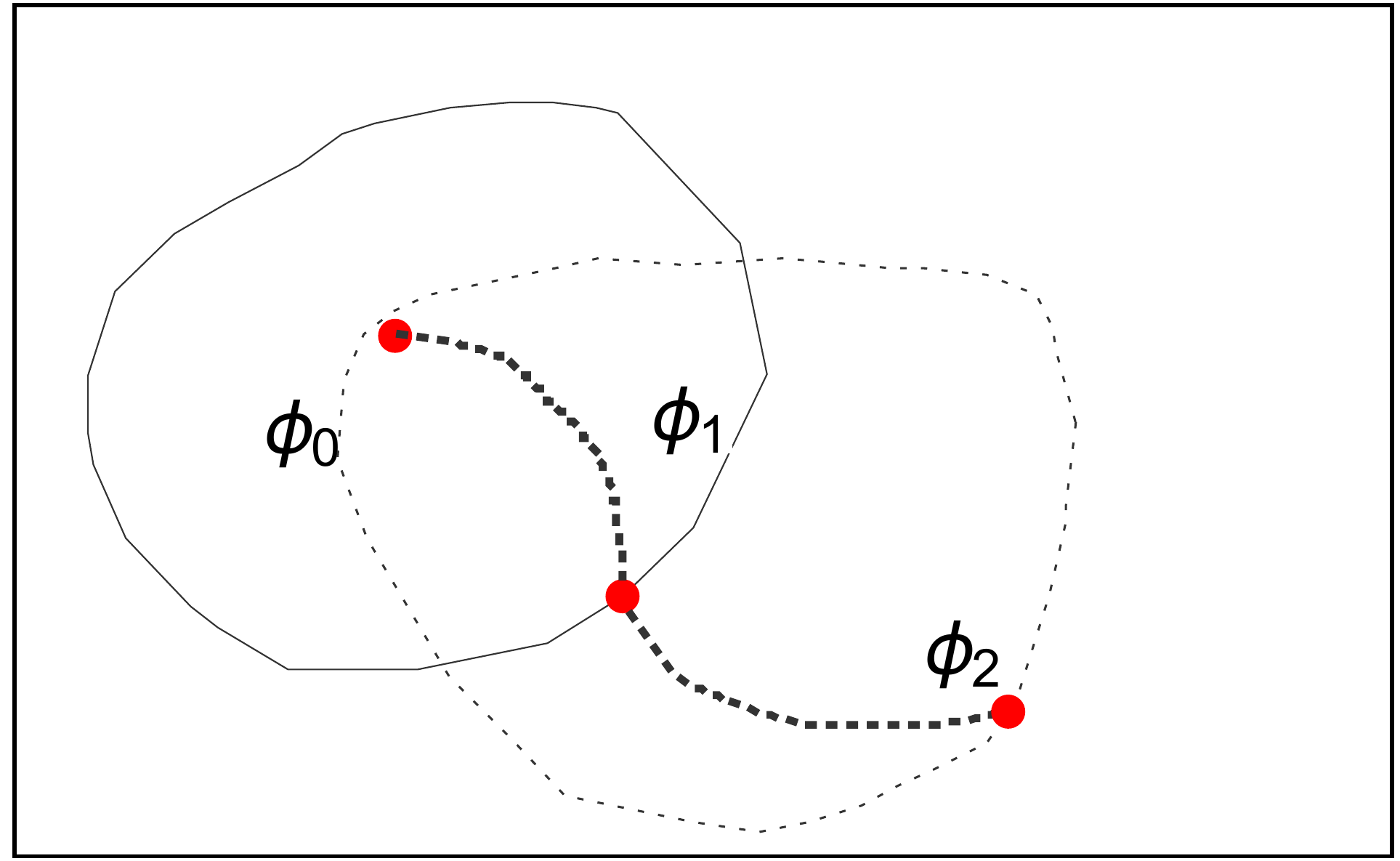} 
   \caption{
Starting from some point ${\bf\phi}_0$ in the field space, the field can be evolved to the point ${\bf\phi}_1$ at the edge of the region of validity of the Taylor expansion of the potential (solid line region).  One can then expand the potential around ${\bf\phi}_1$ and evolve the field to ${\bf\phi}_2$, the edge of validity region for Taylor expansion around ${\bf\phi}_1$ (dotted line region), and so on.
   }
   \label{fig:localExp}
\end{figure}
The Taylor expansion (\ref{Taylor}) gives a good approximation within some radius $r$ in the field space. 
This radius gets larger for higher values of $n$.

Once we know the potential in the vicinity of ${\bf\phi}_0$, 
we can evolve the field from ${\bf\phi}_0$ to some point ${\bf\phi}_1$ near the edge of validity of the Taylor expansion. At this point we want to generate a new approximation for the potential to extend this evolution (see Fig.~\ref{fig:localExp}).  This can be done by finding the probability distribution for the $n$-th derivatives $A^{(1)}_{j_1, \ldots j_n}$ at ${\bf\phi}_1$ under the conditions that $U_0$, $A^{(0)}_{j_1, \ldots j_m} ~ (m=1,\ldots , n)$, $U_1$ and $A^{(1)}_{j_1, \ldots j_m} ~ (m=1,\ldots , n-1)$  have specified values.  For example, if the expansion is up to second order, $n=2$, we find the distribution for the Hessian matrix at ${\bf\phi}_1$ for the given values of the potential, its gradient and the Hessian at ${\bf\phi}_0$ and of the potential and its gradient at ${\bf\phi}_1$.  

What power $n$ should be chosen depends on $N$, the  dimensionality of the field space.  For small $N$
it is less expensive to keep high orders in Taylor expansion, while for large $N$ it is better to keep $n$ small and use many more intermediate points with a smaller radius of validity.  In fact, the optimal strategy in the large $N$ case may be to set $n=1$, except for the first point $\phi_0$, where we need an expansion to second order to check the validity of the slow-roll conditions.

\subsection{Generating random potentials by interpolation}

We note finally that the following method of generating random potentials may be useful in cases when the number of fields $N$ is not very large. 
We specify a set of points $\{{\bf\field}_1, \ldots, {\bf\field}_\kappa\}$ and define random variables $\alpha_j \equiv U({\bf\phi}_j)$.\footnote{The same analysis can be done for specifying derivatives  of arbitrary order at different points using $\expec{\pot_{i_1,\ldots, i_k} (\field_1) \pot_{j_1,\ldots, j_n}(\field_2)}=\partial_{\boldsymbol \field_{1,i_1}}\ldots\partial_{\boldsymbol \field_{1,i_k}}\partial_{\boldsymbol \field_{2,j_1}}\partial_{\boldsymbol \field_{2,j_n}}F\( |\boldsymbol\field_1-\boldsymbol\field_2|\)$.}
The correlators of these variables are given by Eq.\eqref{Correlation},
\bel{Correlations2}
M_{ij}=\expec{\alpha_i, \alpha_j}= F\( |\boldsymbol\field_1-\boldsymbol\field_2|\)~, 
\ee
and the probability distribution for $\alpha_j$ is 
\bel{dist}
	P(\alpha) = \frac{\sqrt{{\rm det} K_{ij}}}{(2\pi)^{\kappa/2}} e^{-\frac12 \alpha_i K_{ij}\alpha_j}~, 
\ee
where $K=M^{-1}$. 

If $\cal O$ is the matrix that diagonalizes $M$, i.e. ${\cal O}^T M {\cal O}= {\rm diag} \{\mu_1, \mu_2, \ldots, \mu_\kappa \}$, the same matrix diagonalizes $K$, ${\cal O}^T K {\cal O}= {\rm diag} \{\lambda_1, \lambda_2, \ldots, \lambda_\kappa \}$, where $\lambda_j = \mu_j^{-1}$.  The probability distribution for the variables $\beta_i={\cal O}_{ij} \alpha_j$ transformed by this matrix is given by 
\bel{dist2}
	P(\beta) = \frac{\sqrt{\prod_i \lambda_i }}{(2\pi)^{\kappa/2}} e^{-\frac12 \lambda_i \beta_i^2}~.
\ee
We can generate the $\beta_i$'s from the set of normal distributions (\ref{dist2}) and then transform them back. The most costly part of generating this sample is diagonalizing $M$, which grows as $\kappa^3$.  Hence, we can generate up to several hundred values of the potentials $U({\bf\phi}_j)$ in a very short time.  This method is not very useful for a large number of fields, as there we need a lattice of many points to capture the statistics of the potential. But for a relatively small number of fields ($N\lesssim 10$) this method is usually cheaper than generating the functions using their Fourier series.

\section{Conclusions}\label{sec:Conclusion}

In this paper we developed analytic and numerical techniques for calculating probability distributions in random Gaussian landscapes.  As an application, we studied the statistics of slow-roll inflation in a one-dimensional landscape where the correlation length $\Lambda$ in the field space is small compared to the Planck scale (so that inflation is small-field).  In such a landscape, inflation occurs either at local maxima (hilltops) or at inflection points of the potential.   In our analytic treatment, apart from the slow roll approximation, we used the "small-field approximation", which assumes that the inflaton potential remains nearly constant during inflation and can therefore be expanded in Taylor series up to cubic order around inflection points or local maxima.  Our analytic results agree very well with numerical simulations, up to relatively large values of the correlation length, such as $\Lambda = 0.5$.

Hilltop inflation is necessarily eternal, while for inflection-point inflation one may be interested in the maximal number of e-folds $N_{\rm tot}$.  We find that the probability distribution for $N_{\rm tot}$ is $P(N_{\rm tot})\propto N_{\rm tot}^{-3}$.  This form of the distributions is not sensitive to the correlation function of the Gaussian ensemble.  

We calculated the probability distributions for the spectral index $n_s$ and its running $\alpha_s$, for both hilltop and inflection-point inflation.  These distributions also have a universal form, independent of the Gaussian correlator.  The observed values of $n_s$ and $\alpha_s$ are in the mid-range of the inflection-point distributions.  In this sense a small-field Gaussian ensemble is consistent with observations.

We should emphasize, however, that the probability distributions we calculated here cannot be directly applied to make predictions in our observable region.  For that one would also need to know the distributions for the initial conditions at the onset of the slow roll and for the properties of the vacuum state at the end of the slow roll.  Inflection-point inflation is likely to begin after tunneling from a metastable vacuum state.  The initial conditions are then determined by the corresponding instanton solution of the Euclidean field equations.  Efficient methods for finding such solutions have recently been developed in Refs.~\cite{AliKen}.  These methods can be combined with the statistical methods we developed here to study the inflationary statistics.  The initial condition issue is also entangled with the measure problem of eternal inflation, which presently remains unresolved.  (For a review of the measure problem see, e.g., \cite{Freivogel}.)

We have outlined how our analysis can be extended to a large number of fields and to large-field inflation with $\Lambda\gg 1$.   An attractive feature of our approach is that it does not suffer from potential inconsistencies inherent in the Dyson Brownian Motion technique, which is the method that has been used in much of the earlier literature.

In Appendix \ref{sec:perturbativity} we addressed the problem raised by Dine in Ref.~\cite{Dine:2015szg}, where he argued that large landscape models may be perturbatively inconsistent.  He estimated quantum corrections to $n$-point interaction vertices and argued that these corrections get unacceptably large in a landscape with a large number of fields $N$.  We attempted a more careful analysis, which suggests that a generic landscape with $N\lesssim 200$ may still be consistent.  Furthermore, we find that axion-type models, with a momentum cutoff scale $M_{\rm cut}\ll \Lambda$ may allow much larger values of $N$.

\section{Acknowledgement}
We are grateful to Richard Easther, Alan Guth, Mark Hertzberg, Ken Olum and Ken Van Tilburg for useful discussions. This work  is supported by National Science Foundation under grant 1518742. M.Y. is supported by the JSPS Research Fellowships for Young Scientists. 
\appendix

\section{Perturbative consistency}\label{sec:perturbativity}

In this Appendix we discuss the perturbative consistency of coupling constants, following Ref.~\cite{Dine:2015szg}. 

We can expand the potential $U(\phi)$ at a given point in the field space, ${\bf\phi} = 0$, 
with the expansion coefficients given by the derivatives of $U$ 
that obey the Gaussian probability distribution. 
In particular, the coefficients of $2n$-point interaction satisfy 
\beq
 \left< U^{(2n)}_{i_1 i_2 \dots i_{2n}} U^{(2n)}_{j_1 j_2 \dots j_{2n}} \right> \sim \frac{U_0^2}{\Lambda^{4n}} \delta_{i_1 i_2 \dots i_{2n} j_1 j_2 \dots j_{2n}}, 
\label{UU}
\eeq
where 
\beq
\delta_{i_1 i_2 \dots i_{2n} j_1 j_2 \dots j_{2n}} = \delta_{i_1 j_1} \dots \delta_{i_{2n} j_{2n}} + {\rm all~permutations}.
\label{delta}
\eeq
and we have used $\sigma_{2n}^2 \sim N^{4n} U_0^2 / \Lambda^{4n}$.
Note that there are $4n$ indices and 
the number of terms on the right hand side of Eq.~(\ref{delta}) is 
\beq
{\cal N}_{4n} = \frac{(4n)! }{ (2n)! 2^{2n}},
\label{calN}
\eeq 
where the factor of $1/(2n)!$ comes from the permutation of $2n$ Kronecker deltas and $1/2^{2n}$ comes from the permutation of $2n$ pairs of indices of the Kronecker deltas.

\begin{figure}[t] 
   \centering
   \includegraphics[width=3.5in]{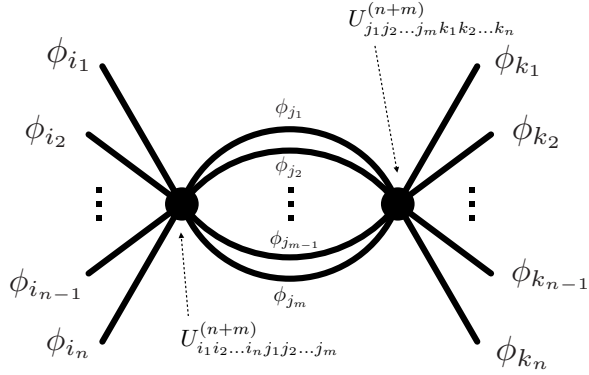} 
   \caption{
   Example of a Feynman diagram that describes a quantum correction to $2n$-point coupling. 
   In this diagram, $m$-pairs of indices are contracted and there are $(m-1)$ loop integrations. 
   }
   \label{fig:diagram}
\end{figure}

The lowest-order quantum correction to the $2n$-point coupling $U^{(2n)}_{i_1 \dots i_n k_1 \dots k_{n}}$ is given by the diagram shown in Fig.~\ref{fig:diagram}, 
where $m$ ($\ge 2$) is the number of internal lines. 
Denoting the cutoff scale as $M_{\rm cut}$, 
this diagram can be estimated as 
\beq
 \delta U^{(2n)}_{i_1 \dots i_n k_1 \dots k_{n}} \sim 
 \frac{M_{\rm cut}^{2m-4} }{(4 \pi)^{2m-2}} 
 \sum_{j_1, j_2, \dots, j_m}
 {U^{(n+m)}_{i_1 i_2 \dots i_n j_{1} j_{2} \dots j_{m}} U^{(n+m)}_{j_1 j_2 \dots j_m k_1 k_2 \dots k_n}}. 
\eeq
where $1/(4 \pi)^{2m-2}$ is the loop factor~\cite{Manohar:1983md}. 

 With the aid of Eq.~(\ref{UU}), we obtain the following estimate for the averaged over the ensemble correction to the $2n$-point vertex:
\beq
  \delta U^{(2n)}_{i_1 \dots i_n k_1 \dots k_{n}} \sim 
 \frac{M_{\rm cut}^{2m-4} }{(4 \pi)^{2m-2}} 
 N^{m} \frac{U_0^2}{\Lambda^{2(n+m)}}
 \delta_{i_1 \dots i_n k_1 \dots k_{n}}.
\label{deltaU}
\eeq
The power of $N$ here is two times larger than that obtained by Dine in Ref.~\cite{Dine:2015szg}.
 The reason is that he assumed that loops of different fields contribute with random signs, while in our Gaussian ensemble all contributions come with the same sign.

The perturbativity breaks down when the quantum correction (\ref{deltaU}) is larger than 
the unperturbed coupling constant, whose typical value can be estimated from the correlation function \eqref{UU} with $j_l = i_l$ for $l = 1,2,3, \dots, 2n$. 
First, let us consider the case when $i_1 = i_{n+1} < i_2 = i_{n+2} < \dots < 
i_{n} = i_{2n}$. 
The typical value of such a coupling constant is given by 
\beq
 U^{(2n)}_{i_1 i_2 \dots i_n i_1 i_2 \dots i_n} 
 &\sim 
 \sqrt{\left< U^{(2n)}_{i_1 i_2 \dots i_n i_1 i_2 \dots i_n} U^{(2n)}_{i_1 i_2 \dots i_n i_1 i_2 \dots i_n} \right>} 
 \nonumber\\
 &\sim \frac{U_0}{\Lambda^{2n}} \sqrt{\delta_{i_1 i_2 \dots i_n i_1 i_2 \dots i_n i_1 i_2 \dots i_n i_1 i_2 \dots i_n}}. 
\eeq
Here, $\delta_{i_1 i_2 \dots i_n i_1 i_2 \dots i_n i_1 i_2 \dots i_n i_1 i_2 \dots i_n}$ has four copies of each index $\{i_1, i_2, \dots, i_n\}$.  For each set of four, we have three possible combinations of the Kronecker deltas, and since we assume all $i_j$ to be different, no other deltas contribute.  Hence we obtain 
\beq
 \delta_{i_1 i_2 \dots i_n i_1 i_2 \dots i_n i_1 i_2 \dots i_n i_1 i_2 \dots i_n} = 3^n, 
\eeq
for $i_1 < i_2 < \dots < i_{n}$. 

We should now compare this tree level coupling with the quantum correction of Eq.~(\ref{deltaU}), 
where $k_l = i_l$ for $l = 1,2,3, \dots, n$. 
Noting that $\delta_{i_1 \dots i_n k_1 \dots k_{n}} = 1$ for $i_1 = k_{1} < i_2 = k_{2} < \dots < i_{n} = k_{n}$, 
we obtain  the following condition for the perturbative stability of the theory:
\beq
\frac{1}{3^{n/2}} \left( \frac{N}{ (4\pi)^2}\frac{M_{\rm cut}^2}{\Lambda^2}\right)^m \frac{(4\pi)^2 U_0}{M_{\rm cut}^4} \lesssim 1.
\label{perturbative consistency}
\eeq

A similar constraint arises in the case when all indices $U^{(2n)}$ are equal to one another. 
In this case, $\delta_{i_1 \dots i_{2n} j_1 \dots j_{2n}} = {\cal N}_{4n}$
in Eq.~(\ref{UU}), because $i _1 = i_2 = \dots = i_{2n} = j_1 = j_2 = \dots = j_{2n}$. 
Thus a typical value of the coupling constant is given by 
\beq
 U^{(2n)}_{i_1 i_2 \dots i_{2n}} \sim 
 \frac{U_0}{\Lambda^{2n}} {\cal N}_{4n}^{1/2}
\eeq
for $i _1 = i_2 = \dots = i_{2n}$. 
Similarly, we have
$\delta_{i_1 \dots i_{n} k_1 \dots k_{n}} = {\cal N}_{2n}=(2n)! / (n! 2^n)$ 
in Eq.~(\ref{deltaU}) with $i _1 = i_2 = \dots = i_{n} = k_1 = k_2 = \dots = k_{n}$. 
Thus we obtain a similar constraint to \eqref{perturbative consistency} when we compare the tree level coupling constant and its quantum correction.

This constraint should be satisfied for all values of $m$.   Thus, we must have
\beq
\frac{N}{ (4\pi)^2}\frac{M_{\rm cut}^2}{\Lambda^2} \lesssim 1.
\eeq
One can expect the cutoff scale to be of the order $M_{\rm cut}\sim\Lambda$.  Then this condition reduces to $N\lesssim (4\pi)^2 \sim 160$, which may still allow a large landscape.  On the other hand, in axion-type models, the cutoff scale may be much smaller than $\Lambda$ and the resulting bound on $N$ much weaker.  For example, in the case of the QCD axion, the role of $\Lambda$ is played by the Peccei-Quinn scale $f_{PQ}$ and the cutoff scale is $\Lambda_{QCD}\ll f_{PQ}$.

\section{Calculating the inverse matrix for third order correlation}\label{sec:InverseMatrix}
Here we show the derivation of $K$, the inverse of the matrix $M$ introduced in \eqref{thirdOrderM}.  Let's assume the inverse matrix has the form 
\bel{inverseMatrix}
	K=\( \begin{array}{cc}
	L \delta_{IJ} &  X \delta_{IJ} \delta_{KL} + W \delta_{IK}\delta_{JL} + O \delta_{IL} \delta_{JK} \\
	X \delta_{IJ} \delta_{KL} + W \delta_{IK}\delta_{JL} + O \delta_{IL} \delta_{JK}& K_{IJK, LMN}
	\end{array}
	 \)
\ee
The most general form of $K_{IJK,LMN}$ is given by
\beal{Kijklmn}
K_{IJK, LMN}&=& C_1  \delta_{IJ}\delta_{KL}\delta_{MN}+C_2 \delta_{IJ}\delta_{KM}\delta_{LN}+ C_3 \delta_{IJ}\delta_{KN}\delta_{ML}\nn
        	   &+&C_4 \delta_{IK}\delta_{JL}\delta_{MN}+C_5 \delta_{IK}\delta_{JM}\delta_{LN}+ C_6 \delta_{IK}\delta_{JN}\delta_{ML}\nn
	          &+&C_7  \delta_{IL}\delta_{JK}\delta_{MN}+C_8 \delta_{IL}\delta_{JM}\delta_{KN}+ C_9 \delta_{IL}\delta_{JN}\delta_{KM}\nn
	 	   &+&C_{10}\delta_{IM}\delta_{JK}\delta_{LN}+C_{11} \delta_{IM}\delta_{JL}\delta_{KN}+ C_{12} \delta_{IM}\delta_{JN}\delta_{KL}\nn
	          &+&C_{13} \delta_{IN}\delta_{KL}\delta_{JM}+C_{14} \delta_{IN}\delta_{KM}\delta_{JL}+ C_{15} \delta_{IN}\delta_{JK}\delta_{LM}~.
\eea

We know the most general form of $ \alpha_{IJ}K_{JL}\alpha_{LN} $ must be rotationally invariant. This forces 
\be
	 \alpha_{IJ}K_{JL}\alpha_{LN}= D_1 \eta_i \eta_i+ D_2 \eta_i \rho_{ijj} + D_3 \rho_{ijk}\rho_{ijk} + D_4 \rho_{iik}\rho_{jjk}, 
\ee
which creates many constraints between the coefficients in \eqref{Kijklmn}.
$K$ is a symmetric matrix and it should remain invariant under $(IJK)\leftrightarrow (LMN)$. This forces $C_1=C_{15}, C_2=C_6, C_4=C_{10}$ and $C_{12}=C_{14}$. We also see that in fact terms 5, 12, 13 and 14 are identical because of the constraints $I\le J \le K$.  They all vanish if two of the indices are different and otherwise are equal.  Therefore, we do not need 12, 13 and 14. We drop all the duplicate terms and this 
 simplifies \eqref{Kijklmn} to
\beal{KijklmnV2}
K_{IJK, LMN}&=& C_1  \delta_{IJ}\delta_{KL}\delta_{MN}+C_2 \delta_{IJ}\delta_{KM}\delta_{LN}+ C_3 \delta_{IJ}\delta_{KN}\delta_{ML}\nn
        	   &+&C_4 \delta_{IK}\delta_{JL}\delta_{MN}+C_5 \delta_{IK}\delta_{JM}\delta_{LN}+ C_2 \delta_{IK}\delta_{JN}\delta_{ML}\nn
	          &+&C_7  \delta_{IL}\delta_{JK}\delta_{MN}+C_8 \delta_{IL}\delta_{JM}\delta_{KN}+ C_9 \delta_{IL}\delta_{JN}\delta_{KM}\nn
	 	   &+&C_4\delta_{IM}\delta_{JK}\delta_{LN}+C_{11} \delta_{IM}\delta_{JL}\delta_{KN}+ C_1\delta_{IN}\delta_{JK}\delta_{LM}~.
\eea
After some algebra one can show 
\beal{totalSum}
	\rho_{IJK}K_{IJK, LMN}\rho_{LMN} &=& 2C_1 \sum_{i\le k\le j} \rho_{iik}\rho_{kjj}+2C_2 \sum_{i\le j}\rho_{iij}\rho_{jjj}+C_3 \sum_{i,j\le k}\rho_{iik}\rho_{jjk} 
	\nn
	&+&2C_4 \sum_{i\le j}\rho_{iii}\rho_{ijj} 
	+C_5\sum_i \rho_{iii}\rho_{iii} +  C_7\sum_{k\le i,j}  \rho_{iik}\rho_{jjk}
	\nn
	&+& C_8 \sum_{i\le j \le k}\rho_{ijk}\rho_{ijk}+ C_9\sum_{i\le j} \rho_{ijj}\rho_{ijj}+ C_{11}\sum_{i\le j} \rho_{iij}\rho_{iij} ~.
\eea
On the other hand, rotational symmetry forces 
\beal{rotInvar}
	\rho_{IJK}K_{IJK, LMN}\rho_{LMN} = M_1 \rho_{ijk}\rho_{ijk}+ M_2 \rho_{iij}\rho_{jkk}~.
\eea

Now let's examine the terms in \eqref{totalSum} closely: First the term containing $C_8$:
\be
	\rho_{IJK}\rho_{IJK}=\frac16 \( \sum_{i,j,k}\rho_{ijk}\rho_{ijk}+ 3 \sum_{i,j}\rho_{iij}\rho_{iij}+2 \sum_{i} \rho_{iii}\rho_{iii}\)~.
\ee
This produces a term which is invariant and has the structure $\rho_{ijk}\rho_{ijk}$ plus other terms. After some computation we can show 
\beq
	\sum_{i,j,k}\rho_{iik}\rho_{jjk} =2\sum_{i\le k\le j} \rho_{iik}\rho_{jjk}+\sum_{i,j\le k} \rho_{iik}\rho_{jjk} +\sum_{i, j\ge k} \rho_{iik}\rho_{jjk}-2\sum_{i,j}\rho_{iii}\rho_{jji}
	- \sum_{i}\rho_{iii}\rho_{iii}~.
	\nonumber\\
	\label{sum1}
\eeq
Comparing \eqref{totalSum}, \eqref{rotInvar} and  \eqref{sum1} it is clear we need $C_1=C_3=C_7$. Combining all we get 
\beal{totalSum2}
	\rho_{IJK}K_{IJK, LMN}\rho_{LMN}&=& \sum_{i,j,k}\[C_1\rho_{iik}\rho_{jjk}+ \frac{C_8}6 \rho_{ijk}\rho_{ijk}\]+2C_1\sum_{i,j}\rho_{iii}\rho_{ijj} 
	\nn
	&+&	\(C_1+\frac{C_8}{3}+C_5\)\sum_i \rho_{iii} \rho_{iii} 
	+  \frac{C_8}{2}\sum_{i,j}\rho_{iij}\rho_{iij}+2C_2 \sum_{i\le j}\rho_{iij}\rho_{jjj}
	\nn
	&+&2C_4 \sum_{i\ge j}\rho_{iij}\rho_{jjj} +    C_9\sum_{i\le j} \rho_{ijj}\rho_{ijj}+C_{11}\sum_{i\ge j} \rho_{ijj}\rho_{ijj} ~.\nn
\eea
This forces $C_2=C_4$ and $C_9=C_{11}$. Putting all these together and rewriting \eqref{totalSum2} we get
\beal{totalSum2}
	\rho_{IJK}K_{IJK,LMN}\rho_{LMN}&=& \sum_{i,j,k}\[C_1\rho_{iik}\rho_{jjk}+ \frac{C_8}6 \rho_{ijk}\rho_{ijk}\] +  \(\frac{C_8}{2}+C_9\)\sum_{i,j}\rho_{iij}\rho_{iij}\nn 
	&+&2(C_1+C_2)\sum_{i,j}\rho_{iii}\rho_{ijj}+	\(C_1+\frac{C_8}{3}+C_5+2C_2+C_9\)\sum_i \rho_{iii} \rho_{iii}~.\nn
\eea
This forces $C_2=-C_1$, $C_9=-\frac12 C_8$ and $C_5= C_1 + C_8/6$. So out of 15 variables in \eqref{KijklmnV2} we only need to determine $C_1$ and $C_8$. Similarly, by demanding $\eta_i \rho_{jkl}$ be invariant under rotation we get 
\bea
	\sum_{i,j\le k\le l} \eta_i K_{i,jkl} \rho_{jkl} &=&  \sum_{i, j\le k\le l }\eta_i \( X \delta_{ij}\delta_{kl}+ W \delta_{ik}\delta_{jl}+  O \delta_{il}\delta_{kj}\) \nn
	&=& X \sum_{i<j} \eta_{i}\rho_{ijj} + (W+X) \sum_i \eta_i \rho_{iii}+ O \sum_{j\le i} \eta_i \rho_{jji}~.
\eea
It is clear that we need $O=X=-W$. Therefore out of 19 unknowns in \eqref{inverseMatrix} only four are left: $X, L, C_1$ and $C_8$. We get these by directly multiplying $K$ and $M$. Before doing this, let's summarize the elements of $K$ after all these simplifications:
\beal{kSummary}
	K_{IJK, LMN}&=& C_1  (\delta_{IJ}\delta_{KL}\delta_{MN}+ \delta_{IJ}\delta_{KN}\delta_{ML} +\delta_{IL}\delta_{JK}\delta_{MN}+\delta_{IN}\delta_{JK}\delta_{LM}+ \delta_{IK}\delta_{JM}\delta_{LN}\nn
			&&- \delta_{IJ}\delta_{KM}\delta_{LN}- \delta_{IK}\delta_{JL}\delta_{MN}-\delta_{IM}\delta_{JK}\delta_{LN}- \delta_{IK}\delta_{JN}\delta_{ML})\nn
        	      &+&\frac16 C_8 (6\delta_{IL}\delta_{JM}\delta_{KN}-3 \delta_{IL}\delta_{JN}\delta_{KM}- 3\delta_{IM}\delta_{JL}\delta_{KN}+\delta_{IK}\delta_{JM}\delta_{LN})~, \nn
	      K_{I,JKL}&=&X ( \delta_{IJ} \delta_{KL} - \delta_{IK}\delta_{JL} +  \delta_{IL} \delta_{JK})~.
\eea

By multiplying the matrices $M$ and $K$ we should get the unit matrix, which translates into 
\bea
	&&B L + (N+2)AX=-1~, \nn
	&&(N+4)X Y - LA =0~, \nn
	&& B X + A C_1 (N+2)+\frac12 A C_8=0~, \nn
	&& Y C_8=1~.
\eea
Solving these equations gives 
\bea
	C_1 &=& -\frac{A^2+B Y}{2 Y \left(A^2 (N+2)+B (N+4) Y\right)}\nn
	C_8&=& \frac1Y~, \nn
	L&=&-\frac{(N+4) Y}{A^2 (N+2)+B (N+4) Y} ~,\nn
	X&=& -\frac{A}{A^2 (N+2)+B (N+4) Y}~.
\eea

\section{Single-valuedness of the potential}\label{sec:singleValued}

In this Appendix we show that the methods we developed in this paper generate single-valued potentials.  
We also show that, in contrast to the methods used in \cite{Freivogel:2016kxc}, the derivative  of our potentials is well-defined for arbitrarily short distances.  In addition, we demonstrate both analytically and numerically the extent to which Taylor expansions are good representatives of the random Gaussian potentials. 

Let us first consider the values of the potential at two distant points: 
$\pot({\bf\field}_0)=\pot_0$ and $\pot({\bf\field}_1)=\pot_1$. 
Without loss of generality, using translational and rotational invariance of the distribution, we can reduce the problem to a one-dimensional case  with 
$\field= |{\bf\field}_1-{\bf\field}_0|$. Using 
\be
\expec{\pot_0\pot_0}=\expec{\pot_1 \pot_1}= F(0)~, \qquad \expec{\pot_0\pot_1}= F(\field)~,
\ee
the probability distribution for $\pot_0$ and $\pot_1$ can be written as 
\beq
P(\pot_0, \pot_1)&&= \sqrt{\frac{F(0)}{ 2 \pi (F(0)^2-F(\field)^2)} } \exp\[-\frac{F(0)}{2 \left(F(0) ^2-F(\field) ^2\right)} \left(\pot_1-\frac{F(\phi) }{F(0)}\pot_0\right)^2\]
\nonumber\\
&&\qquad  \times
\frac{1}{\sqrt{2 \pi F(0)}} 
\exp\[-\frac{\pot_0^2}{2 F(0)} \].
\label{P(U_1, U_0)}
\eeq
Hence, the distribution of $\pot_1$ is peaked around $\pot_0$ and its width shrinks as $\sqrt{F(0)-F(\field)}$ for $\phi\to 0$. This demonstrates the single-valuedness of the potential in our method.

Let us now 
consider the value of the potential and its first $n$ derivatives at a point $\phi_0$, namely $U(\phi_0), U'(\phi_0), \ldots, U^{(n)}(\phi_0)$. The truncated Taylor expansion would be given by 
\bel{TaylorTest}
    \tilde{U}(\phi)= \sum_{j=0}^n \frac1{j!}U^{(j)}(\phi_0)(\phi-\phi_0)^j~.
\ee
Then we can ask whether or not 
the value of potential $\tilde{U} (\phi)$ ($\phi \ne \phi_0$) is consistent 
with the one indicated by $P( U(\phi_0), U'(\phi_0), \ldots, U^{(n)}(\phi_0), U(\phi))$. 
Quantitatively, we can calculate 
\beq
 \frac{P( U(\phi_0), U'(\phi_0), \ldots, U^{(n)}(\phi_0), U(\phi))}{
P( U(\phi_0), U'(\phi_0), \ldots, U^{(n)}(\phi_0))}, 
\eeq
in the same way as \eqref{P(U_1, U_0)} and compare its $1 \sigma$ region with $\tilde{U} (\phi)$. 
This allows us to check the consistency of using Taylor series. 

\begin{figure}[t] 
   \centering
   \includegraphics[width=2.7in]{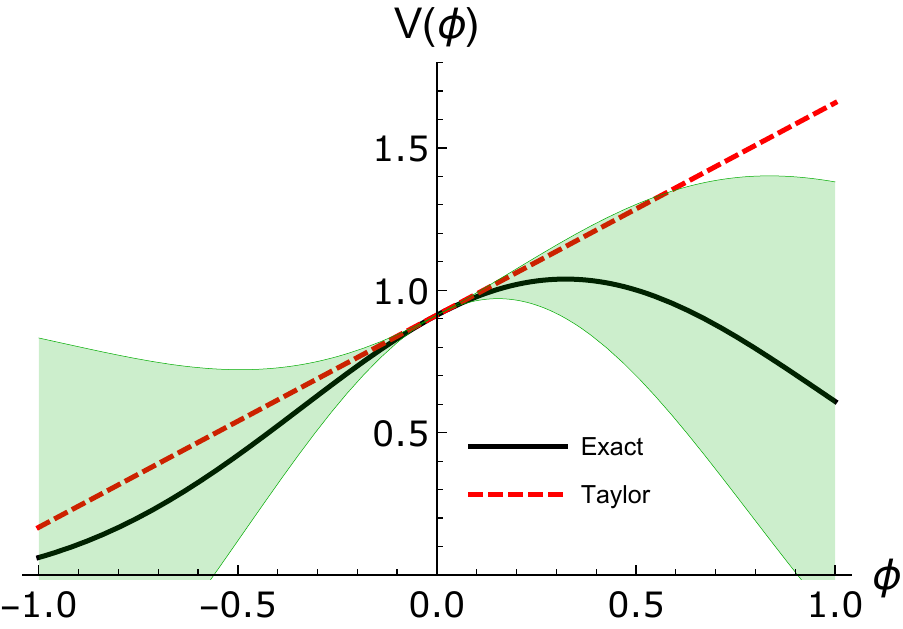} 
   \includegraphics[width=2.7in]{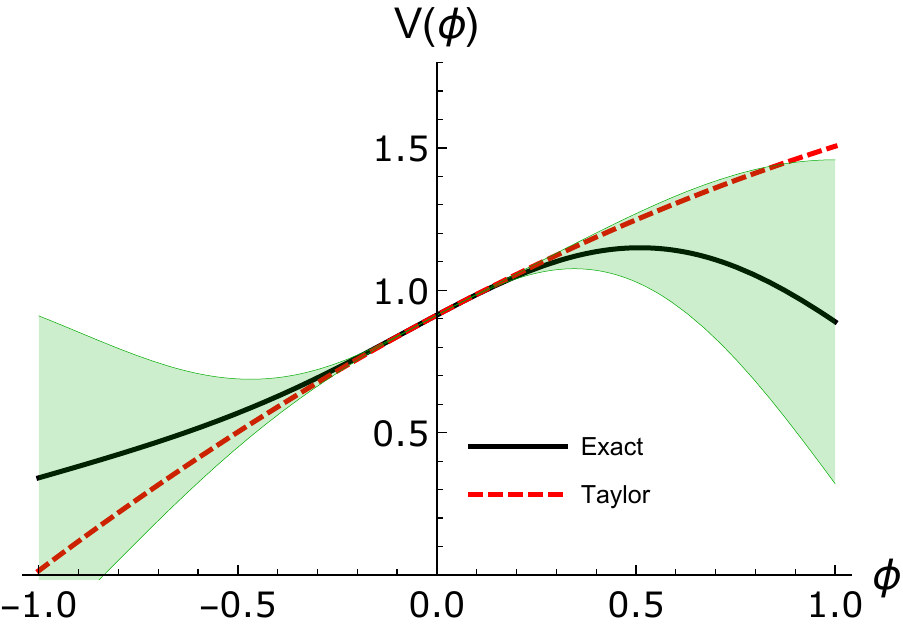}
   \includegraphics[width=2.7in]{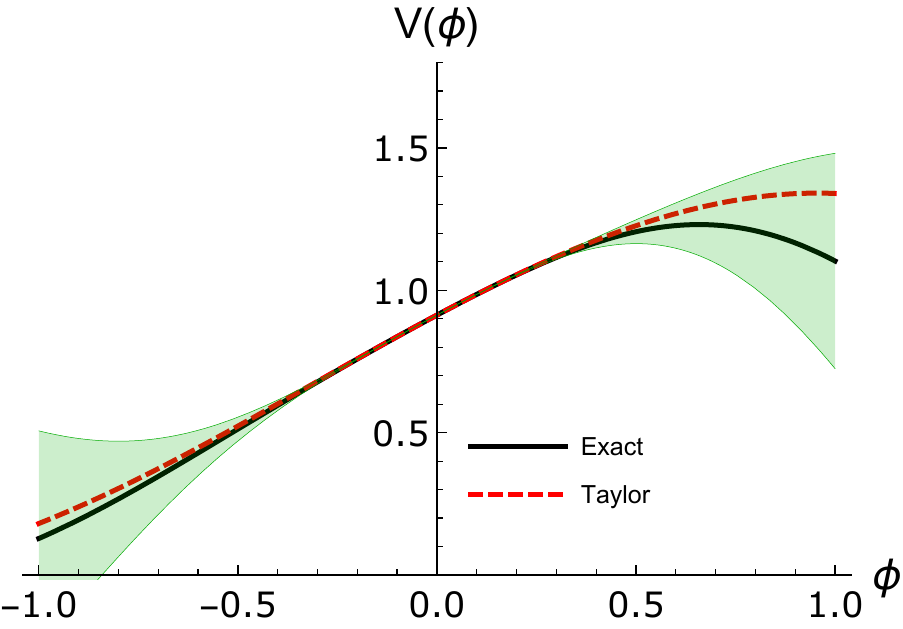} 
   \includegraphics[width=2.7in]{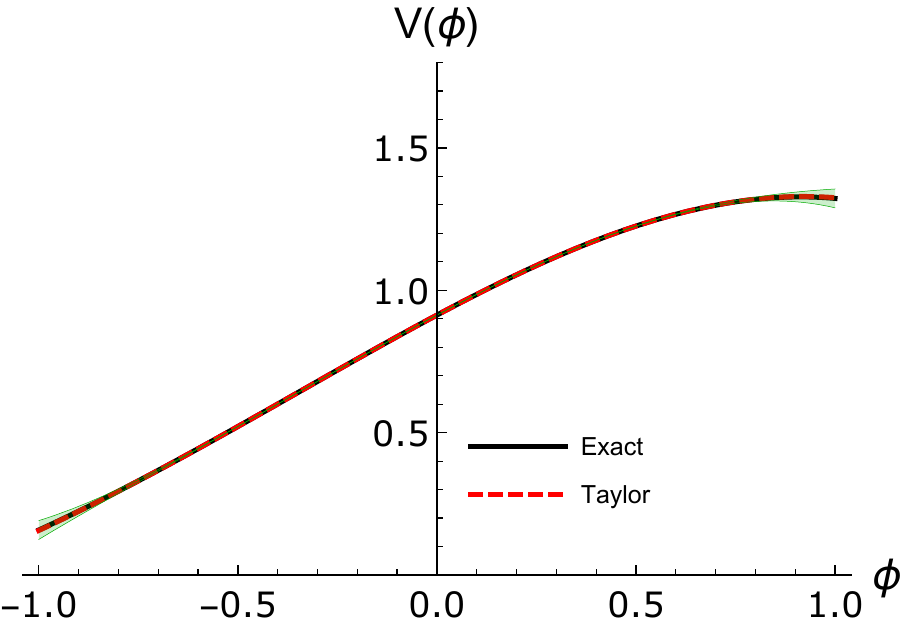}  
   \caption{From upper left to lower right panels we show the ensemble average and the width of the probability distribution for a  potential $U(\phi)$ with Taylor expansion coefficients of order 1,2,3
 and 7 specified at $\phi=0$.  We use the correlation function given in \eqref{gaussianPower} with $U_0=\Lambda=1$. 
The values of the potential and its first four derivative at $\phi=0$ were chosen as $\{0.913,0.745,-0.305,-0.994,-0.506,0.581,0.980,0.219\}$. Changing these priors does not substantially change the results.}
   \label{fig:TaylorCompare}
\end{figure}

As an illustration we chose a power spectrum given by 
\bel{gaussianPower}
\expec{U(\phi_1), U(\phi_2)}= U_0^2 e^{-|\phi_1-\phi_2|^2/\Lambda^2},
\ee
with $U_0=\Lambda=1$. We computed the ensemble average of $U(\phi)$ and its width for Taylor expansions of different orders.  An example with expansions to orders $n=1,2,3,7$ is shown in Fig.~\ref{fig:TaylorCompare}.  We see that the width of the distribution decreases rapidly with $n$.  In this and other examples that we considered, a cubic approximation appears to work  well for half the correlation length. These conclusions are not sensitive to the priors chosen at $\field_0$.

\end{document}